\DocumentMetadata{}
\documentclass[manuscript, screen]{acmart}

\authorsaddresses{}

\AtBeginDocument{%
  }

\DeclareUnicodeCharacter{FF0C}{,}

\definecolor{activegold}{RGB}{255,193,61}
\definecolor{lightorange}{RGB}{230, 170, 50}
\definecolor{lightgreen}{RGB}{121,210,121}
\definecolor{lightteal}{RGB}{121,199,210}
\definecolor{lightblue}{RGB}{100,212,239}
\definecolor{lightpurple}{RGB}{153,102,255}
\definecolor{deepgreen}{RGB}{0, 100, 0}

\usepackage{xspace}
\usepackage{multirow}

\newcommand{\pzh}[1]{{\color{black} #1}}
\newcommand{\fhx}[1]{{\color{black} #1}}
\newcommand{\peng}[1]{{\color{black} #1}}
\newcommand{\penguin}[1]{{\color{black} #1}}
\newcommand{\shiwei}[1]{{\color{black} #1}}
\newcommand{\zqw}[1]{\color{black} #1}
\newcommand{\ReviseZQW}[1]{\color{black} #1}
\newcommand{\Original}[1]{\color{black} #1}
\newcommand{\ReviseFinal}[1]{\color{black} #1}

\newcommand{\name}{{\textit{AskArt}}}

\newcommand{\ie}{\textit{i.e.,} }
\newcommand{\eg}{\textit{e.g.,} }

\setcopyright{acmlicensed}
\copyrightyear{2025}
\acmYear{2025}
\acmDOI{XXXXXXX.XXXXXXX}

\acmConference[CSCW'25]{In Proceedings of The 2025 ACM
SIGCHI Conference on Computer-Supported Cooperative Work \& Social Computing (Conference CSCW ’25)}{October 2025}{Bergen, Norway}

\begin{document}

\title{Exploring the Usage of Generative AI for Group Project-Based Offline Art Courses in Elementary Schools}

\author{Zhiqing Wang}
\authornote{Corresponding author}
\email{wangzhq83@mail.sysu.edu.cn}
\affiliation{%
  \institution{Sun Yat-sen University}
  \city{Guangzhou}
  \state{Guangdong}
  \country{China}
}

\author{Haoxiang Fan}
\email{pengzhh29@mail.sysu.edu.cn}
\affiliation{%
  \institution{Sun Yat-Sen University}
  \city{Guangzhou}
  \country{China}
}

\author{Shiwei Wu}
\email{wushw28@mail2.sysu.edu.cn}
\affiliation{%
  \institution{Sun Yat-Sen University}
  \city{Guangzhou}
  \country{China}
}

\author{Qiaoyi Chen}
\email{chenqy99@mail2.sysu.edu.cn}
\affiliation{%
  \institution{Sun Yat-Sen University}
  \city{Guangzhou}
  \country{China}
}

\author{Yongqi Liang}
\email{shiyanxuexiaoqiqi@126.com}
\affiliation{%
  \institution{Xiangzhou Experimental School of Zhuhai}
  \city{Zhuhai}
  \country{China}
}

\author{Zhenhui Peng}
\authornote{Principle Investigator.}
\email{pengzhh29@mail.sysu.edu.cn}
\affiliation{%
  \institution{Sun Yat-Sen University}
  \city{Guangzhou}
  \country{China}
}

\renewcommand{\shortauthors}{Zhiqing et al.}

\begin{abstract}
The integration of Generative Artificial Intelligence (GenAI) in K-6 project-based art courses presents both opportunities and challenges for enhancing creativity， engagement, and group collaboration. 
This study introduces a four-phase field study, involving in total two experienced K-6 art teachers and 132 students in eight offline course sessions, to investigate the usage and impact of GenAI. 
Specifically, based on findings in Phases 1 and 2, we developed \name{}, an interactive interface that combines DALL-E and GPT and is tailored to support elementary school students in their art projects, and deployed it in Phases 3 and 4. 
Our findings revealed the benefits of GenAI in providing background information, inspirations, and personalized guidance. However, challenges in query formulation for generating expected content were also observed. Moreover, students employed varied collaboration strategies, and teachers noted increased engagement alongside concerns regarding misuse and interface suitability. This study offers insights into the effective integration of GenAI in elementary education, presents \name{} as a practical tool, and provides recommendations for educators and researchers to enhance project-based learning with GenAI technologies.
% The integration of Generative Artificial Intelligence (GenAIs) in K-6 project-based art courses presents both opportunities and challenges for enhancing creativity， engagement, and group collaboration. This study introduces a four-phase field exploration involving two experienced K-6 art teachers and 132 students, utilizing eight sessions to investigate the impact of GenAIs. We developed and deployed \name{}, an interactive interface combining DALL-E and GPT, tailored to support elementary school students in their art projects. Our findings revealed the benefits of GenAIs in providing background information, personalized guidance, and visual representations. However, challenges in query formulation and input management were also observed. Moreover, Students employed varied collaboration strategies, and teachers noted increased engagement alongside concerns regarding misuse and interface suitability. This study offers insights into the effective integration of GenAIs in elementary education, presents \name{} as a practical tool, and provides recommendations for educators and researchers to enhance project-based learning with GenAI technologies.
\end{abstract}



% \received{2 July 2024}
% \received[revised]{29 October 2024}
% \received[revised]{10 December 2024}
% \received[accepted]{4 March 2025}

\maketitle

\section{Introduction}

Art courses in elementary schools or K-6 education \footnote{In this paper, we use K-6, primary schools, and elementary schools interchangeably, which refer to the initial stage of formal education and typically serve children from around ages 6 to 11 or 12.} are popular around the world. 
In these courses, instructors implement a variety of teaching methods with activities (\eg group or individual projects, discussions, exhibitions) to foster children's creativity, critical thinking, self-identity, and social competencies \cite{gadsden2008arts,garvis2011breaking}. 
Project-based learning (PBL) is one such method that is extensively applied in art education \cite{lu2022project, zhang2023cultivating, polina2022student, al2023creating}. 
This approach engages students in real-world projects and often encourages them to discuss ideas and work together in peer groups. 
The effectiveness of PBL has been well-documented {\ReviseFinal{in}} fostering students' creativity, autonomy, collaboration, problem-solving skills, and reflective thinking in real-world contexts \cite{chiang2016effect, johnson2013everyday, hawari2020project,samaniego2024creative}. 

\penguin{
From the teachers' perspectives, a typical PBL process in art course can include planning (\eg topic, question, deadlines), testing (\eg project progress, facilitation), and reflecting (\eg presentation, explanation) \cite{hawari2020project}. 
This paper focuses on the testing phase of the PLB process, in which students in a small group need to work together to finish a task given by the teachers. 
{\ReviseFinal{During this phase, students gather relevant information, brainstorm ideas, and implement solutions. However, elementary school students may struggle with these tasks due to limited knowledge, teamwork skills, and self-directed learning abilities \cite{chen2021forms, miao2024project}. }}
% In this phase, students often seek related information about the project and engage in ideation and implementation of the solution, which, however, could be challenging for elementary school students due to the lack of knowledge and skills in teamwork and self-directed learning \cite{chen2021forms, miao2024project}. 
In offline classroom where there is normally one art teacher in one class, students can query the teacher for support when they face challenges during information seeking and solution ideation. 
Nevertheless, the teacher's support may not be in time, especially when managing multiple groups with varying needs and abilities \cite{zhuge2009teaching,wildermoth2012project, miao2024project}. 
}

\penguin{
Recent advances in Generative AIs (GenAIs) offer a promising alternative to offering in-situ information seeking and ideation support to students in their PBL art courses. 
For example, the text-to-image generation techniques like DALL-E enable the production of high-quality images from textual descriptions \cite{ramesh2021zero,esser2024scaling} and have been used in supporting ideation in various design-related tasks, \eg facilitating creativity \cite{designingwithai},  explore design possibilities\cite{designprompt}, and enhancing collaborative ideation\cite{texttoimagecatalyst}.
The conversational agents powered by large language models (LLMs, \eg GPT) can answer any user query in nearly real-time, which could bridge knowledge gaps and promote creativity \cite{decisionmaking, boscardin2024chatgpt, dwivedi2021artificial, pardos2023oatutor}. 
Prior studies have underscored the role of GenAIs in various educational roles, such as adaptive learning, collaborative learning, and AI-assisted content creation \cite{pardos2023oatutor,sharples2023towards}. 
Despite the potentials, little research has investigated the usage of GenAIs to support students in information seeking and ideation in their group-based offline art courses in elementary schools. 
Applying GenAIs in this education scenario is unique against using them in higher education \cite{zhou2020designing} or individual art learning activities \cite{karpouzis2024tailoring}. 
For one thing, many students in elementary schools may find the GenAIs non-transparent regarding their capabilities and implications \cite{van2023emerging}. 
For another, it is not always feasible, or often not recommended for group projects in offline classrooms, to offer each elementary school student a device for accessing GenAI in art courses. 
Besides, it is unknown how teachers would perceive and collaborate with GenAI to facilitate students in the PBL art courses. 
Correspondingly, three research questions are under-explored: 
\begin{itemize}
    \item \textbf{RQ1}: How would elementary students use GenAIs, \ie the text-to-image generation techniques and large language models in our case, for information seeking and solution ideation in their group projects in offline art courses? 
    \item \textbf{RQ2}: What interaction design with GenAIs would be helpful for supporting students in PBL art courses? 
    \item \textbf{RQ3}: How would GenAIs affect the the process of the PBL art course from teachers' perspectives?
\end{itemize}
Answering these questions can contribute to the CSCW communities with empirical findings and implications of facilitating elementary students in group-based information-seeking and ideation activities with GenAIs. 
% Therefore, there is a need to explore how GenAIs can be adopted in group project-based IBL and how students would perceive and interact with them in their art projects.
}

\begin{figure}[h]
  \centering
  \includegraphics[width=\linewidth]{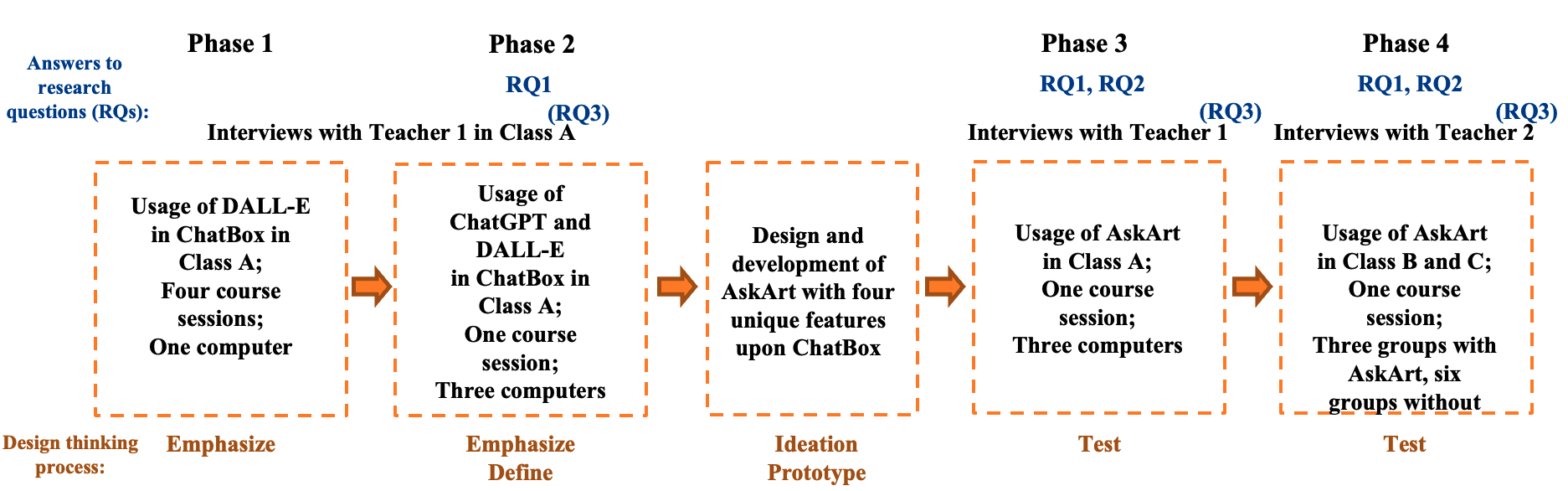}
  \caption{
  \penguin{Overview of our four-phases study as well as its relationship to the design thinking process and three research questions, \ie RQ1 about elementary students' usage of GenAIs in their PBL art courses, RQ2 about helpful interaction design with GenAIs for supporting them, and RQ3 about teachers' thoughts on the impact of GenAIs on their courses.}
  % Overview of four phases study, mapped to 3 RQs. Phase 1 and Phase 2 emphasize understanding user needs and defining design requirements through interviews with Teacher 1 and observing the usage of GenAI tools in Class A. Phase 3 focuses on ideating and prototyping AskArt, integrating unique features based on insights from previous phases. Phase 4 tests AskArt in real classroom settings with Class B and Class C, comparing three groups using AskArt with five groups using a baseline condition. RQs addressed in each phase are indicated in blue.
  }
  \label{fig:4phase}
  \Description{}
\end{figure}

To this end, we collaborated with two experienced K-6 art teachers in China to run a four-phase field study \penguin{(Fig. \ref{fig:4phase})} to explore the usage of GenAIs in a total of eight-course sessions. % (six 1-hour sessions and two 40-minute sessions). 
In Phase 1, we introduced DALL-E, a text-to-image model, to the \textbf{Class A} with 24 students who form three 8-person groups in their project-based art course. 
In four course sessions, whenever they wish, students in Class A can use DALL-E in the Chatbox \footnote{\url{web.chatboxai.app}} interface via one computer in the classroom. 
In Phase 2, we additionally introduced GPT, an LLM-powered conversational agent, to Class A for a one-course session. 
\penguin{Results of Phases 1 and 2 answer RQ1 by highlighting DALL-E's values in inspiring ideas and GPT's benefits in providing background information and implementation guidance. 
We also found that students use GPT and DALL-E in a sequential order in Phase 2 but often struggled with query formulation to get the expected content.} 
\penguin{These findings informed our design of \name{}, an interface that aims to answer RQ2 by integrating GPT and DALL-E with tailored features to support K-6 students in art courses. }
% Based on the feedback from students and Teacher 1 of Class A, we developed \name{}, an interactive interface that integrates GPT and DALL-E with tailored features to support K-6 students in art courses. 
Specifically, \penguin{compared to the traditional interface in ChatBox,} \name{} provides: i) Project-Related Introduction of GenAI; ii) Audio-to-Text Integration; iii) Suggested Follow-Ups; iv) \penguin{Select and Generate}. % Auto-Generated DALL-E Prompts; v) 
% \peng{[describe the key unique features here]}. 

\penguin{To evaluate \name{}, in Phase 3, we deployed it to class A for another course session to have qualitative comparisons between \name{} and Chatbox. 
In Phase 4, we further compared \name{} to a baseline condition without access to any computer devices via a co-designed one-course session with Teacher 2 for Class B and Class C, in which only three randomly selected groups in each class can access \name{}. 
Results from Phases 3 and 4 {\ReviseFinal{provide additional insights into RQ1 by demonstrating GenAI’s role in supporting}} information seeking and ideation in PBL art courses. {\ReviseFinal{They also inform}} RQ2 by highlighting the benefits of \name{}'s ``Suggested Follow-Ups'' and ``Select and Add Content in GPT as the prompts'' to DALL-E  for fostering elementary students' interaction with GenAIs. 
As for RQ3, students with \name{} in Phase 4 reported significantly {\ReviseFinal{higher}} satisfaction than those without \name{} in the PBL art course. 
They developed different strategies for integrating GenAIs into their group work, balancing between collective brainstorming and individual tasks. 
Across the four phases, the two teachers noted in the interviews that GenAIs increased students' engagement in the courses but concerned the misuse of GenAIs in the group-based art projects. 
Based on our findings, we discuss opportunities and implications for using GenAIs to help elementary students seek information and ideate solutions during group project-based courses. 
}

\pzh{
Our contributions to the CSCW communities are three-fold. 
First, we offer first-hand findings on the usage of GenAIs in group project-based K-6 art courses in offline classrooms. 
The findings showcase the benefits, challenges, usage patterns, and impact of GenAIs in these courses. 
Second, we present an interactive interface \name{} that customizes the interaction with GenAI for K-6 students \penguin{to facilitate information seeking and solution ideation} in their group project-based art courses. 
Third, we offer actionable insights for educators and researchers to leverage GenAIs in \penguin{elementary students'} group project-based learning \penguin{and more broadly, group-based information-seeking and ideation activities}. 
}
% By analyzing these aspects, this paper aims to provide a critical evaluation of GenAI in enhancing PBL within K-6 art education from the perspectives of both teachers and students, and to establish a set of design requirements based on real-world interactions between K-6 students and GenAI in PBL art classes. The findings will contribute to the development of design principles for future educational GenAI-powered systems and offer actionable insights for educators and developers aiming to leverage GenAI in PBL and K-6 art education.

\penguin{
In the following sections, we first review related work on PBL, GenAIs for supporting art activities, and GenAIs for education. Then, we present the overview of our four-phases studies (\autoref{sec:overview_four_phase}), followed by the details and findings in Phases 1 and 2 (\autoref{sec:phase_1_2}), the design of \name{} (\autoref{sec:askart_design}), and studies in Phases 3 and 4 (\autoref{sec:phase_3_4}). 
Lastly, we discuss implications of our work in \autoref{sec:discussion}. 
}

\zqw{
\section{Related Work}
\subsection{\pzh{Project-Based Learning} in Art Education}
Project-based learning (PBL) represents a shift from traditional teacher-centered instruction to a more student-focused approach, where learning occurs within a specific context, students engage actively, and knowledge is shared through social interactions \cite{kokotsaki2016project}. This method requires students to address real-world problems through extended inquiry, fostering active and personalized learning experiences \cite{chiang2016effect, tang2023taking}. PBL aims to provide students with a broad skill set that transcends traditional educational methods, enhancing the practical aspects of learning and boosting employability \cite{belwal2020project}. The effectiveness of PBL has been validated in \pzh{various} practice \cite{miao2024project, usmeldi2019effect, hawari2019challenges, wuntu2022implementation}. 
%For instance, a study in Slovakian secondary schools compared the outcomes of teaching economics through PBL with traditional methods \cite{maros2023project}. In this study, 123 students were split into control and experimental groups, with the control group receiving standard verbal-visual instruction and the experimental group participating in PBL. Both groups were instructed by the same teacher and covered identical material over the same period.The findings demonstrated that students in the experimental group performed better academically, confirming the greater effectiveness of PBL compared to traditional teaching methods.
\fhx{\pzh{For example}, a study in Slovakian secondary schools compared the outcomes of teaching economics through PBL with traditional methods \cite{maros2023project}. It involved 123 students divided into control and experimental groups, with the latter participating in PBL. The findings showed that the experimental group performed better academically, confirming the effectiveness of PBL over traditional teaching methods.}

PBL has also shown considerable promise in art education by engaging students in meaningful learning activities. In progressive education, particularly in the arts, emphasis is placed on hands-on activities and projects that encourage exploration and concept discovery through active involvement \cite{pavlou2024learning}. This aligns with the principles of PBL. Implementing PBL in art education has been confirmed to boost students' motivation, as students learn by doing and applying their ideas to complete directed projects, leading to more effective learning outcomes \cite{mccullough2018implementing}.

However, PBL in art education presents challenges from both student and teacher perspectives. Designing and preparing projects can be time-consuming for teachers \cite{zhuge2009teaching, chen2021forms}, and projects must be carefully designed to meet educational goals 
\fhx{to keep students engaged \cite{wildermoth2012project}. } 
% while keeping students engaged \cite{wildermoth2012project}. 
Students often struggle to take the initiative in the learning process, where they must identify problems and apply their knowledge to find practical solutions \cite{ahern2010case, bledsoe2012concept, lutsenko2018case}. 
\fhx{Continuous guidance from teachers and facilitators is recommended to improve students’ PBL skills, particularly in teamwork and self-directed learning \cite{chen2021forms}. However, this adds to the already substantial workload reported by teachers involved in PBL \cite{miao2024project}.}

%Continuous guidance from teachers and facilitators on teamwork and self-directed learning is recommended to improve students’ PBL skills \cite{chen2021forms}, although this adds to the already substantial workload reported by teachers involved in PBL \cite{miao2024project}.

\pzh{Our work is motivated by the benefits of PBL in art education and aims to mitigate these challenges in implementing group project-based art courses in elementary school students. 
\peng{
Compared to the PBL practices mentioned above, our target scenario could be more challenging, as elementary school students usually are less experienced in independent project completion than senior students and the group PBL with two or more students would be more dynamic than the individual PBL.}
}

% \subsection{GenAI supporting art projects}
\pzh{\subsection{Supporting Art Activities with GenAI}}
With the growing availability of GenAI technologies, including large language models (LLMs) and image generators, researchers are exploring opportunities for using GenAI to assist in art creation. 
Generative AI models have proven to be highly effective for art creation, particularly in image generation. 
\pzh{For example,} \citet{ghosh2019interactive} introduced a GAN-based interactive image generation system designed to create complete images from partial user strokes or sketches.
This system also functions as a recommender, assisting users throughout their creative process to help them produce their desired images. With  the introduction of DALL-E 3, MidJourney, and Stable Diffusion, high-quality AI image generation became available to the public and was actively involved in the creation process of artworks.
\pzh{For instance,} \citet{Braintax} introduced BrainFax, a tool that integrated with Stable Diffusion to help humans generate and edit images. Professional designers participated in the evaluation for BrainFax shared their enthusiasm and saw potential or use in their own work. 

In addition to image generators, recent work has begun to explore how LLMs were involved in projects in terms of ideation \cite{kim2023effect, tholander2023design}. %, as LLMs provides. 
For example, \citet{shaer2024ai} explored the integration of LLMs into the creative process, focusing on both the divergence stage of idea generation and the convergence stage of idea evaluation and selection. 
They developed a collaborative group-AI Brainwriting ideation framework that incorporates an LLM to enhance the group ideation process. 
Their study evaluated the idea-generation process and the resulting solution space, demonstrating that integrating LLMs in Brainwriting can significantly enhance both the ideation process and its outcomes.
The integration of multiple GenAI models can further offer comprehensive support for the art project process. 
\pzh{For example}, \citet{wander} utilized LLMs and Stable Diffusion to exemplify the comprehensive support of GenAI in facilitating art projects through ideation, problem-solving, and visual support. In their work, LLMs were used for generating narrative travelogues based on the location information, while Stable Diffusion was used to create the images of future aligning the travelogues and the photo of that specific location.  This comprehensive approach illustrates how GenAI can support the entire artistic process, from ideation and storytelling to visual representation, thereby enhancing the overall creative workflow. 

\pzh{The successful usage of GenAI in art activities inspires us to leverage generative technology to support students in their group art projects. 
We contribute to this line of related work with first-hand findings and insights about how students perceive and interact with GenAI in the group project-based art courses.}

\pzh{
\subsection{GenAI for Education}}
Extensive research has explored the application of GenAI to support \pzh{students' learning activities from various aspects} \cite{sharples2023towards}. %various aspects of the student learning process 
For example, \citet{codeaid} developed CodeAid, an LLM-powered programming assistant that provides timely and personalized feedback without revealing direct code solutions. CodeAid assists students by answering conceptual questions, generating pseudo-code with explanations, and annotating incorrect code with fix suggestions. Deployed in a 700-student programming class over a 12-week semester, CodeAid demonstrated how GenAI can offer immediate and customized guidance, which enhanced the learning experience by promoting cognitive engagement and providing valuable support tailored to individual student needs. Similarly, Khan Academy has begun deploying ``Khanmigo''~\cite{chen2023integrating}, based on GPT, to offer one-on-one tutoring for learners of all backgrounds and skill levels and to provide guided lesson planning for educators. 
In addition to personalized guidance, GenAI has been leveraged to support educational environments by fostering essential soft skills like teamwork and collaboration. For example, \citet{Genaiforgroup} introduce the ``CoBi'' system, a multi-party AI partner designed to enhance students' collaborative skills by focusing on the relationship dimension of collaboration. Unlike traditional AI-driven tutoring tools that emphasize one-on-one instruction and personalized learning, CoBi assists students in co-negotiating classroom agreements along four dimensions: respect, equity, community, and thinking. By employing advanced speech and language technologies, CoBi monitors small group collaborative discourse for evidence of these agreements and visualizes the findings to motivate reflection. These feedback visualizations enable students to understand their strengths in developing collaborative relationships, identify areas for improvement, and cultivate critical AI literacy skills.

While GenAI tools are more accessible to higher education students, lower education students remain an important group of users that cannot be neglected \cite{emeragingtechnologies}. The UNICEF (United Nations Children's Fund) has developed guidelines for ``AI and child rights policy'' with international experts, emphasizing that ``the rights of children, as current users of AI-enabled systems and the future inhabitants of a more AI-saturated world, must be a critical consideration in AI development'' \cite{mathiyazhagan2023children}. 
Much HCI research has been devoted to studying the integration of GenAI in lower education settings (\eg junior or senior high school). However, AI in elementary school education has received relatively little attention. 
\citet{zhou2020designing} organized past findings of integrating GenAI into lower education and proposed a conceptual framework to support researchers, designers, and educators in creating effective AI learning experiences for K-12 students.
%Despite these efforts, as summarized in Maarten et al.'s work \cite{emeragingtechnologies}, the education of emerging technologies, including AI, primarily focuses on middle school students, with few records specifically addressing elementary students. 
\fhx{\citet{van2023emerging} summarize that despite various efforts, the education of emerging technologies, including AI, primarily focuses on middle school students, with few records specifically addressing elementary students.}
Although limited, attempts to integrate AI in elementary school education have proven effective and beneficial. 
% For instance, ChatScratch, powered by Stable Diffusion, ControlNet, and LLMs, facilitates project planning, asset creation, and programming, thereby promoting autonomous learning of Scratch for elementary school children. 
\fhx{For example, ChatScratch \cite{chen2024chatscratch}, which uses Stable Diffusion with ControlNet and LLMs, helps with project planning, asset creation, and programming. This tool promotes autonomous learning of Scratch for elementary school children.}

However, among these few attempts, none have integrated GenAI into existing curricula for elementary school students within a real classroom setting. 
\pzh{
% Overall, while we understand the effectiveness and challenges of project-based learning (PBL) in art education for elementary school students and the potential of GenAI to support the PBL learning process, m
Much less is known about how elementary school children perceive and use GenAI in real practice and how teachers react to the integration of GenAI in a real classroom setting. 
Answering these questions is the aim of our work and can shed lights into equipping offline classrooms with GenAI.
}

}

% \penguin{
% \section{Phases 1 and 2:  Usage of DALL-E and GPT via a Traditional Conversational Interface}
% }
\penguin{\section{Overview of Four-Phase Field Study}
\label{sec:overview_four_phase}}
To \penguin{explore the elementary students' usage of Generative AIs (GenAIs) (RQ1), identify rooms for improvements of GenAIs' interface (RQ2), and examine the impacts of GenAIs (RQ3) in group project-based art courses}, we collaborate with two elementary school teachers (noted as Teacher 1 and Teacher 2) in a local city in China to conduct a four-phase field study in their offline classrooms (Fig. \ref{fig:4phase}). 
\peng{While the settings of authentic project-based art classroom may make it difficult for conducting a rigorous controlled experiment due to noisy factors like students' background and teaching schedules, such settings can help us ensure the ecological validity of our study and gain insightful qualitative results.} 
No alterations were made to the class environment or curriculum content, aside from the inclusion of the GenAIs interface. 
Given the autonomous nature of PBL classes, and in consultation with the teacher, students' usage of the GenAIs interface in the course was entirely voluntary. 
While students were encouraged to engage with GenAIs, no compulsion was applied. 

\begin{figure}[h]
  \centering
  \includegraphics[width=\linewidth]{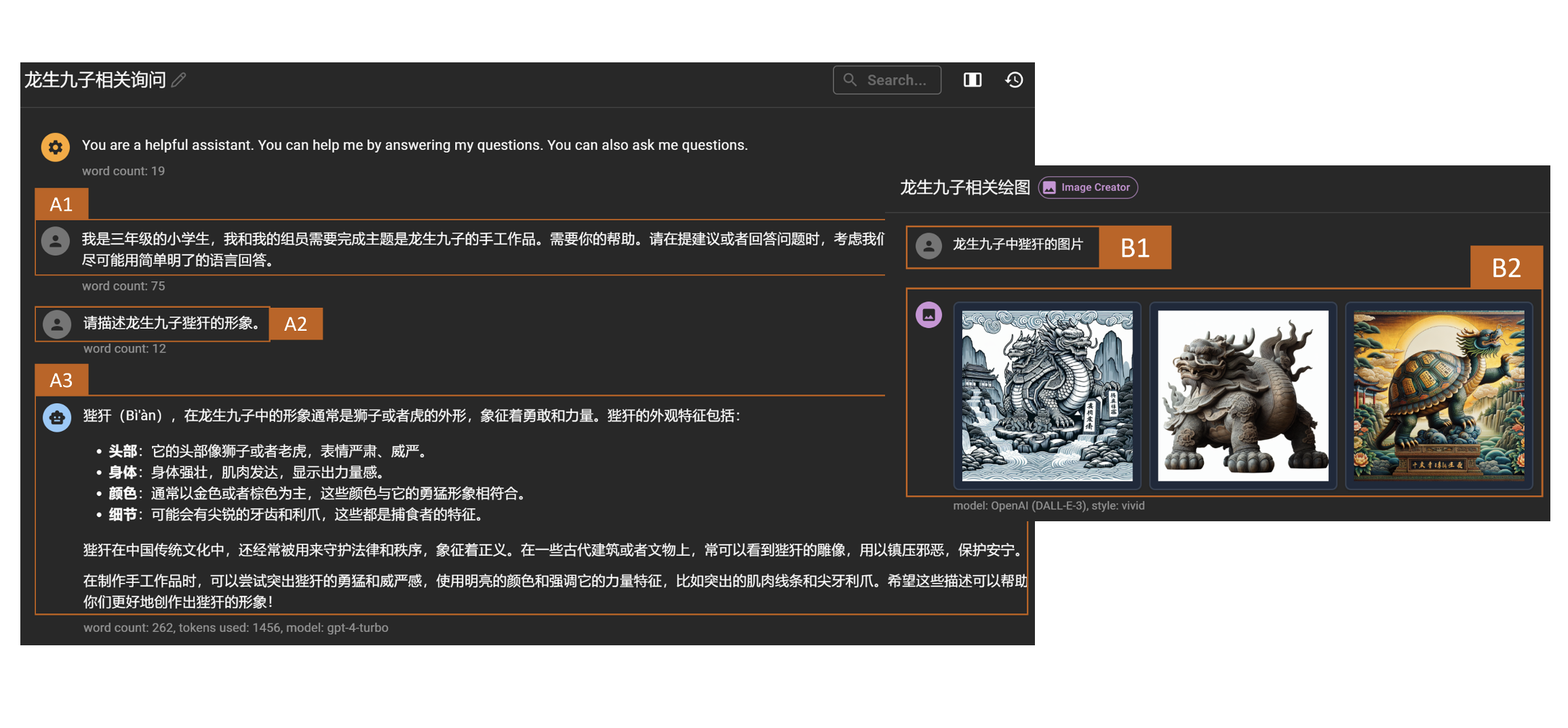}
  \caption{The Chatbox interfaces used in Phases 1 and 2. The left portion of the figure illustrates the question-asking interface and the right one illustrates the image-generation interface.}
  \label{fig:chatbox}
  \Description{This Chatbox interface integrates GPT and DALL-E for comprehensive student support. The left section (A) displays interaction with GPT: students input queries (A1, A2) and receive detailed text responses (A3). The right section (B) shows the DALL-E image generation: students can request (B1) and view images (B2) based on their queries. }
\end{figure}

\penguin{
We followed the design thinking process to plan the four-phase study and adjusted it based on students' and teachers' feedback. 
In the ``Empathize'' and ``Define'' stages, we conducted interviews with Teacher 1 to understand the practices and challenges of enacting project-based learning (PBL) in the art courses and identify what and how appropriate GenAI techniques can help. 
Specifically, Phase 1 serves as a warm-up phase that helps the teacher and students learn about the usage of GenAIs. 
As both Teacher 1 and our research team agreed that the image generator could support students in ideating and picturing the project outcomes in the art course, we started with \textbf{DALL-E} \footnote{\url{https://openai.com/index/dall-e-3/}}, which ``understands significantly more nuance and detail than the previous DALL-E systems and allows users to easily translate their ideas into exceptionally accurate images''. 
Phase 1 lasts for four course sessions, after which the research team and Teacher 1 met and discussed the learned insights. 
Followed Teacher 1's suggestions, in Phase 2, apart from DALL-E, we allowed students to use GPT-4 Turbo, to examine whether and how it could help students seeking information and guidance related to the project. 
In Phases 1 and 2, students in each group can access the GenAIs via a traditional conversational interface \textbf{Chatbox} \footnote{\url{https://web.chatboxai.app/}} (\autoref{fig:chatbox}). 
In the GPT-4 window (\autoref{fig:chatbox} A), students can start with an initial prompt introducing the context (A1), pose questions (A2), and receive detailed text responses (A3). 
In the DALL-E window (\autoref{fig:chatbox} B),  students can submit image queries (B1) and view the generated images (B2).
With the design goals derived from the findings of Phases 1 and 2, in the ``Ideation'' stage of the design thinking process, our research team and Teacher 1 brainstormed interaction features of an interface upon Chatbox that might improve students' efficiency in using GenAIs in their art projects. 
In the ``Prototype'' stage, we developed the \name{} prototype. 
Lastly, in the ``Test'' stage, we conducted one user study in Phase 3 and another study in Phase 4 to evaluate \name{}.
}

\autoref{tab:Phases Overview} summarizes the main setups of the four-phase study. 
Specifically, Teacher 1 has 10 years of art education experience and two years of experience in applying group project-based learning (PBL) in her class. We conduct Phases 1, 2, and 3 in Teacher 1's class (Class A) with 24 students (4 males and 20 females, Grade 3 - 5). 
Students in this class were divided into three 8-person groups during the project-based art course. 
\peng{
Teacher 2 has 20 years of art
education experience and 8 years of experience in applying group PBL in her art courses.} 
% We conducted Phase 4 in two classes of Teacher 2: Class B has 54 males and females (Grade 6), and Class C has 54 males and females (Grade 6). 
\fhx{We conducted Phase 4 in the two classes taught by Teacher 2: Class B with 54 students (Grade 6), and Class C with 54 students (Grade 6).}

% There are four phases in our study, we will now give details on each phase (3.1 - 3.2, 3.5, 3.6),  findings based on phase 1-2 (3.3), and design of AskArt (3.4). 

% \begin{table*}
%   \caption{ Overview of Real Class Experiments Across Four Phases}
%   \label{tab:PBLtasks}
%   \centering
%   \begin{tabular}{ccccc}
%     \toprule
%     Phase&Participants&Duration&Interface&GenAI Tool(s)&Data Collection\\
%     \midrule
%     Phase 1 & Teacher 1 with Class A & 4 classes & Chatbox & DALL-E 3&\\ 
%     Phase 2 & Teacher 1 with Class A & 1 class & Chatbox & DALL-E 3 and GPT-4 Turbo\\ 
%     Phase 3 & Teacher 1 with Class A  & 1 class & AskArt & DALL-E 3 and GPT-4o\\ 
%     Phase 4 & Teacher 2 with Class B and Class C & 1 class & AskArt & DALL-E 3 and GPT-4o\\ 
%     \bottomrule
%   \end{tabular}
% \end{table*}
% \usepackage{multirow}
% \usepackage{multirow}
% \usepackage{multirow}

% \usepackage{multirow}
\begin{table}[]
\caption{Overview of the settings, respondents to the questionnaire, and usage GenAI interface in our four-phases study in the art course sessions.}
  \label{tab:Phases Overview}
\begin{tabular}{ccccccc}
\hline
\multirow{2}{*}{Phase} & \multirow{2}{*}{Participants}                                                                                                              & \multirow{2}{*}{\# Sessions}                                                              & \multicolumn{2}{c}{Questionnaire}                                                                                               & \multicolumn{2}{c}{GenAI}                                                                              \\ \cline{4-7} 
                       &                                                                                                                                            &                                                                                           & No.                                      & Respondents                                                                          & Interface                & Models                                                                      \\ \hline
1                      & \multirow{3}{*}{\begin{tabular}[c]{@{}c@{}}Teacher 1\\ with Class A: \\ three 8-students \\ groups\end{tabular}}                           & 4                                                                                         & \multicolumn{2}{c}{None}                                                                                                        & \multirow{2}{*}{Chatbox} & DALL-E 3                                                                    \\ \cline{1-1} \cline{3-5} \cline{7-7} 
2                      &                                                                                                                                            & 1                                                                                         & 2                                        & P2-1 to P2-6                                                                         &                          & \begin{tabular}[c]{@{}c@{}}DALL-E 3,\\ GPT-4\end{tabular}                   \\ \cline{1-1} \cline{3-7} 
3                      &                                                                                                                                            & 1                                                                                         & 3                                        & P3-1 to P3-5                                                                         & \multirow{5}{*}{AskArt}  & \multirow{5}{*}{\begin{tabular}[c]{@{}c@{}}DALL-E 3,\\ GPT-4o\end{tabular}} \\ \cline{1-5}
\multirow{4}{*}{4}     & \multirow{4}{*}{\begin{tabular}[c]{@{}c@{}}Teacher 2\\ with Class B/C:\\  three/five 6-students groups \\ with/without AskArt\end{tabular}} & \multirow{4}{*}{\begin{tabular}[c]{@{}c@{}}1 in\\ Class B.\\ 1 in\\ Class C\end{tabular}} & \multirow{2}{*}{4}                       & \multirow{2}{*}{\begin{tabular}[c]{@{}c@{}}P4-1 to P4-13\\ with AskArt\end{tabular}} &                          &                                                                             \\
                       &                                                                                                                                            &                                                                                           &                                          &                                                                                      &                          &                                                                             \\ \cline{4-5}
                       &                                                                                                                                            &                                                                                           & \multicolumn{1}{l}{\multirow{2}{*}{4-all}} & \multicolumn{1}{l}{\multirow{2}{*}{P4-1 to P4-108}}                                  &                          &                                                                             \\
                       &                                                                                                                                            &                                                                                           & \multicolumn{1}{l}{}                     & \multicolumn{1}{l}{}                                                                 &                          &                                                                             \\ \hline
\end{tabular}
\end{table}

% \begin{table*}[]
%   \caption{Overview of Real Class Experiments Across Four Phases}
%   \label{tab:Phases Overview}
% \begin{tabular}{lllllll}
% \hline
% \small Phase   & \small Participants           & \small Duration  & \small Interface &\small GenAI Tool(s)       &\small Questionnaire         &\small Respondents \\ \hline
% \tiny Phase 1 & \tiny Teacher 1 with Class A & \tiny 4 classes & \tiny Chatbox   & \tiny DALL-E 3            & \tiny None                  & \tiny None                        \\
% \tiny Phase 2 &             & \tiny 1 class & \tiny Chatbox & \tiny DALL-E 3 and GPT-4  & \tiny Questionnaire 2 & \tiny P2-1 to P2-6  \\
% \tiny Phase 3 &  & \tiny 1 class   & \tiny AskArt    & \tiny DALL-E 3 and GPT-4o &\tiny  Questionnaire 3       &\tiny  P3-1 to P3-5                \\
% \tiny Phase 4 & \tiny Teacher 2 with Class B and Class C & \tiny 1 class & \tiny AskArt  & \tiny DALL-E 3 and GPT-4o      & \tiny Questionnaire 4 & \tiny P4-1 to P4-13 \\
%         &                        &           &           &                     & \tiny Questionnaire 4-tTest & \tiny P4-1 to P4-108              \\ \hline
% \end{tabular}
% \end{table*}

\penguin{
\section{Phases 1 and 2: Usage of DALL-E and GPT via a Traditional Conversational Interface} \label{sec:phase_1_2} 
}

\penguin{
 \subsection{Procedure, Project Task, and Collected Data}
}
\penguin{Prior to the first course session in \textbf{Phase 1}, the research team interviewed Teacher 1, discussing} the most appropriate GenAI techniques that can be introduced to the art course, determining the course sessions in which students can access the GenAIs, and deciding how students can interact with the GenAIs. 
\penguin{As described in \autoref{sec:overview_four_phase}, we agreed that students can use DALL-E in the Chatbox interface in Phase 1.} 
\penguin{Teacher 1 suggested that we could allow students to access GenAIs in the course sessions for ideation and implementation of an artwork for the main project in that semester. 
The students' task in this project is} designing and creating a handcrafted artwork representing one of the dragons from a Chinese tale ``The Nine Sons of the Dragon'', whose names are Qiu Niu, Ya Zi, Chao Feng, Pu Lao, Suan Ni, Bi Xi, Bi An, Fu Xi, and Chi Wen. 
Each of the three groups are assigned with three sons of the dragon to be considered in their projects. 
\penguin{The task prompt for each group is: ``Please design and make a handcraft to show [Name of the Dragon's Son]. Think about its special features and what it means in the ancient stories and Chinese culture.''
} 

% \penguin{[To be continued]}.

% The theme of the PBL art course was to design and create a handcrafted artwork representing one of the dragons from the Nine Sons of the Dragon. 
% For this phase, we utilized DALL-E 3 as the GenAI model, integrated with an interface developed by Chatbox.
% The procedure of Phase 1 is as follows: 
% \begin{itemize}
%     \item \textbf{Interview with Teacher 1} before the first course session. The research team and Teacher 1 discuss the most appropriate GenAI model that can be introduced to the art course, determine the course sessions in which students can access to the GenAI, decide how students can interact with the GenAI. 
%     \item \textbf{Running the four course sessions}. In the first session, Teacher 1 introduces DALL-E to the students via a short tutorial prepared by the research team, which shows several example use cases of DALL-E. As they wish, students can interact with the DALL-E in Chatbox deployed on only one computer. 
%     \item \textbf{Interview with Teacher 1} after the fourth course session. The research team and Teacher 1 discuss \peng{insights and lessons learned from the four sessions and the plans for the coming course sessions}. 
% \end{itemize}

% \textbf{Running the four course sessions}. 
In the first session \penguin{of Phase 1, which is the Week 3 of the 2024 Spring  semester}, Teacher 1 introduced DALL-E to students through a short tutorial prepared by the research team, which shows several example use cases of DALL-E. 
As they wish, students can interact with DALL-E in Chatbox deployed on one single computer \penguin{in the classroom, which is a typical device setting in the elementary schools in main cities in China }. 
\penguin{Phase 1 lasted for four continuous course sessions.}
% \textbf{Interview with Teacher 1} 
At the end of the last session, the research team and Teacher 1 discussed \peng{insights and lessons learned from the four sessions and the plans for the coming course sessions}. 
\penguin{We collected audio recordings of the interviews with Teacher 1.} 

% The workflow began with a pre-interview with the teacher to discuss the most appropriate GenAI model to introduce first. 
% During this interview, we also focused on detailed design aspects, such as determining the phase of the course to introduce the tool, deciding how students would interact with the tool as a group, and whether to provide hands-on guidance. 
% Additionally, we discussed the necessary facilities and the duration of this phase.

% Following the pre-interview, we introduced DALL-E to the students. 
% An initial tutorial was provided during the first class session. 
% Subsequently, one classroom computer was utilized by the entire class, with students interacting with the DALL-E bot independently. 
% This phase encompassed four class sessions.
% \pzh{
% The \textbf{data collected in Phase 1} include: 
% \begin{itemize}
%     \item Audio recordings of the interviews with Teacher 1. 
%     \item Prompts used by students in the four-course sessions. 
% \end{itemize}
% Data collection involved interview recordings with the teacher, including both pre-interviews and post-interviews, to gather insights on the teacher's perspective regarding the students' interactions with DALL-E and its impact. 
% We also saved the chat history between students and DALL-E, documenting how the students engaged with the DALL-E bot during the sessions. 

% \subsection{Phase 2} 
In \textbf{Phase 2,} apart from DALL-E, we also allow students to use GPT-4 Turbo in a separate chat window of Chatbox, as Teacher 1 suggests at the end of Phase 1 that a large language model may help students collect information and search guidance for completing a project.  
% In this phase, the models utilized were DALL-E 3 and GPT-4 Turbo, integrated into Chatbox, which differed from Phase 1 that only used DALL-E 3. 
Phase 2 lasts for one \peng{40-minute} course session with the same project theme in Phase 1.
\penguin{Before the course session, we first interviewed Teacher 1, introducing} the usage of GPT to Teacher 1 and discussing the coordination of the research team to run the course session with GPT and DALL-E. 
Then, three members of our research team came to the classroom. They first introduced the GPT to students with its example use cases. Each of the three groups is assigned one laptop and one member of our team as the teaching assistant (TA). The TA was only responsible for typing the prompts into GPT or DALL-E if necessary but did not directly participate in the group project. 
\penguin{At the end of the course session, we distributed Questionnaire 2} to six randomly selected students, two from each group, and conducted quick interviews with them to make sense of their responses to the questionnaires. 
\penguin{Questionnaire 2 asks: i) the needs and challenges during the information-seeking process before using GenAIs (\ie GPT and DALL-E) in this course session; ii) ratings for GenAIs' assistance in the ideation process (1 - strongly disagree, 7 - strongly agree); and iii) benefits and challenges of using GenAIs.}
\penguin{After the course session, we also interviewed Teacher 1 about} the students' perceptions of GPT and DALL-E, the impact of GenAIs on her class so far, the challenges encountered, and the plan for future course sessions. 
Data collected in Phase 2 include: audio recordings of the interviews with Teacher 1 and students, students' prompts to GPT and those for DALL-E in the course session, responses of six students on Questionnaire 2, and screen recordings of the GenAIs' usage in the computers (with the audio input and camera on) for the three student groups.

% \penguin{[To be continued]}
\penguin{
\subsection{Analysis and Results in Phases 1 and 2}
\subsubsection{Data Analysis} \label{sec:data_analysis} 
\textbf{Interviews}.}
For the audio recordings of interviews with teachers and students collected in \penguin{Phases 1 and 2}, we employed the reflex thematic analysis approach. 
Two researchers independently reviewed the transcripts and responses to identify key themes. 
These themes were synthesized into affinity diagrams and discussed with all authors to ensure consistency and accuracy. 
The outstanding themes from the interview with teachers answer the \textbf{RQ3} and are presented \penguin{later} in \autoref{sec:rq3}, while those from the interview with students help to answer \textbf{RQ1} and are detailed in \penguin{this subsection}. 

% Interviews with teachers and students were conducted in different phases to gather qualitative data. Interviews with teachers took place in phases 1-4, contributing to \textbf{RQ3} by exploring teachers' opinions. Interviews with students were conducted in phases 2-4 to supplement their questionnaire responses, contributing to \textbf{RQ1} regarding students' perceptions of GenAI usage in PBL. Thematic analysis was employed for these interview recordings, with two researchers independently reviewing the transcripts and responses to identify key themes. These themes were synthesized into affinity diagrams and discussed with all authors to ensure consistency and accuracy.

\textbf{Students' Responses to Questionnaires}. 
The collected students' responses to the questionnaires distributed in \penguin{Phase 2} contribute to \textbf{RQ1}. 
To ensure the authenticity of the responses, we took several approach and filter unsuitable responses if applicable. For \penguin{Questionnaire 2}, if a student's response indicated a possible misunderstanding of the question, we reconfirmed their response during follow-up interviews. 
% In Questionnaire 4, each teaching assistant closely reviewed each respondent's questionnaire before collecting it, ensuring data accuracy.
% For Questionnaire 4-all, we observed that some students left comments under the rating questions. If there was a mismatch between the attitudes revealed in the comment and the rating score (\eg, a negative comment for a question rated 7 out of 7), these responses were filtered out, resulting in the exclusion of 9 out of 108 questionnaire sheets.
The qualitative responses (\ie students' comments) to the questions in \penguin{Questionnaires 2} are analyzed via a reflex thematic analysis similar to that for the interview data mentioned above. 

\textbf{Prompts to GPT and DALL-E and Screen Recordings}. 
The logged prompts used by students and the screen course recordings in \penguin{Phase 2} are used to identify patterns and understand students' engagement with the tools, contributing to \penguin{RQ1}. %\textbf{RQ2} (\autoref{sec:rq2}). 
Usage analytics were performed on the logged prompts. Content analysis was conducted to examine the screen recordings, coding the behaviors and interactions observed in the videos. Specifically, students' responses to the answers generated by GenAIs in Phase 2 were coded by two authors and classified into three categories: (1) perceived as correct and helpful, (2) perceived as correct but not helpful, and (3) perceived as incorrect.

\begin{figure}[h]
  \centering
  \includegraphics[width=\linewidth]{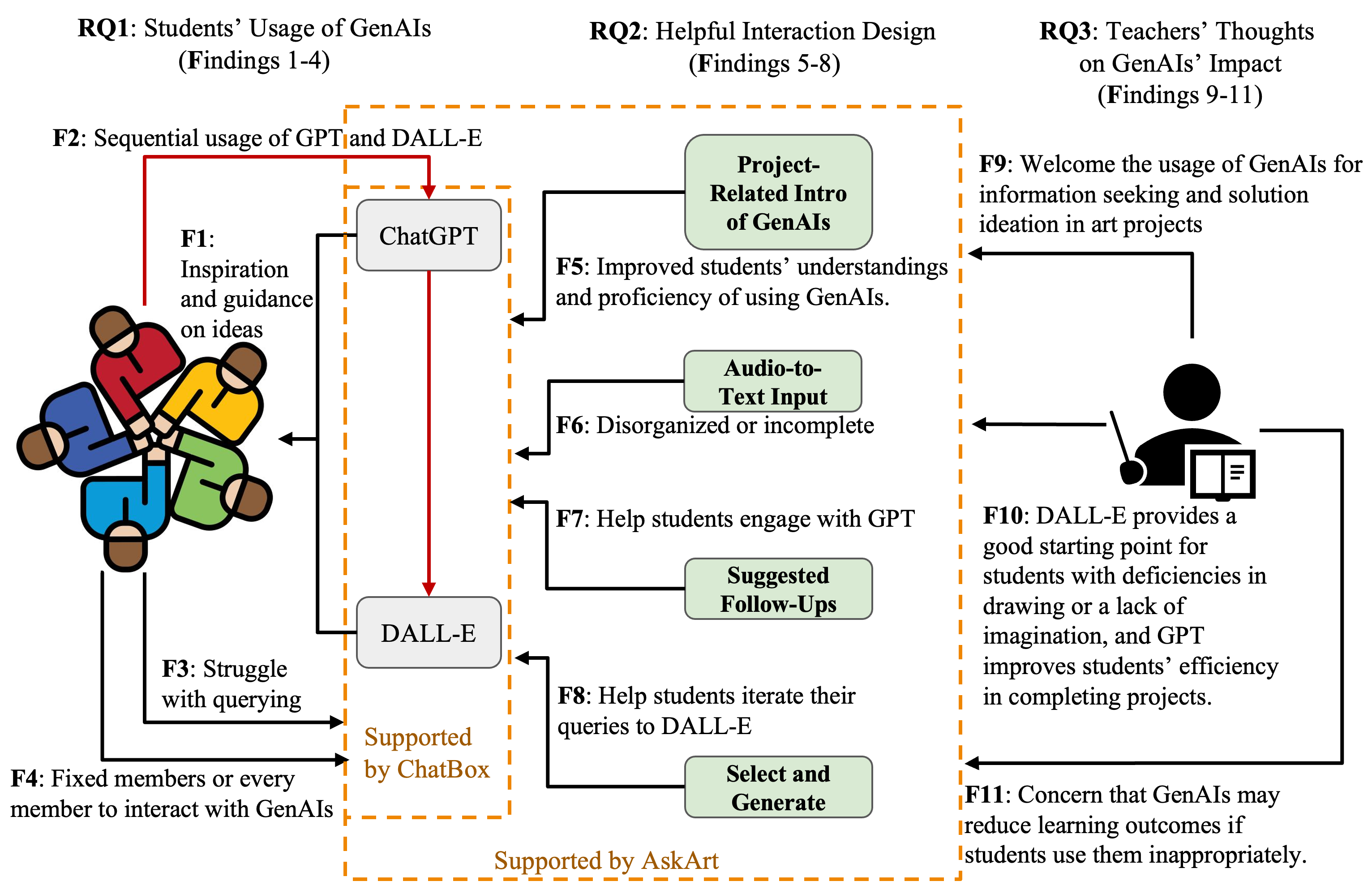}
  \caption{
  \penguin{Summary of our study \textbf{F}indings about elementary students' usage of GenAIs (F1-4 for RQ1), helpful interaction design (F5-8 for RQ2) explored via \name{}, and teachers' thoughts on GenAIs' impact (F9-11 for RQ3).}
  % Overview of key findings organized by(RQs). Findings related to RQ1 illustrate students' usage of GenAIs in group art projects, including their sequential use of GPT and DALL-E (F2), inspiration and guidance on ideas (F1), struggles with query formulation (F3), and collaboration strategies (F4). Findings for RQ2 focus on helpful interaction design, such as the Project-Related Introduction of GenAIs (F5), Audio-to-Text Input (F6), Suggested Follow-Ups (F7), and the Select and Generate feature (F8). Findings for RQ3 reflect teachers’ perspectives, highlighting the benefits of GenAIs in supporting information seeking and ideation (F9), the advantages of DALL-E for students with drawing or imagination deficiencies and GPT for efficiency (F10), and concerns about potential misuse of GenAIs impacting learning outcomes (F11). The system features supported by ChatBox and AskArt are annotated within the diagram.
  }
  \label{fig:summary_of_findings}
  \Description{}
\end{figure}

\penguin{
\subsubsection{Findings in Phases 1 and 2 for RQ1}\label{sec:finding_phase_1_2}
The data analysis revealed students' behavioral patterns, perceived benefits, and encountered challenges when interacting with GenAIs in their group projects, which help to answer our RQ1 posted in the Introduction (Fig. \ref{fig:summary_of_findings}), \ie 
``\textbf{RQ1}: How would elementary students use GenAIs, \ie the text-to-image generation techniques and large language models in our case, for information seeking and solution ideation in their group projects in offline art courses?''. 
}

% \textbf{Sequential Use of GPT and DALL-E}
\penguin{\textbf{Finding 1: DALL-E and GPT inspired students' project ideas and guided the idea implementation, and GPT additionally provided background information about the project}. 
}
In Phase 2, we asked the six respondents to rate the level of their agreement with the statement ``The generated content helps our group come up with more ideas'' and got an average score of $5.67 (SD = 0.82)$, \penguin{indicating that the GenAIs was helpful for solution ideation in the elementary students' art projects}; 1 - strongly disagree, 7 - strongly agree. 
For example, during Phase 2, as observed by the TA, students in Group 2 initially had no idea about the design of the dragon’s eyes. 
After prompting DALL-E with ``images of Qiu Niu from the Nine Sons of the Dragon'', they were inspired by the blue eyes of the dragon in the generated images.
\penguin{Besides, four out of six respondents to Questionnaire 2 in Phase 2 inform their challenges and needs for GenAIs' guidance in implementing their project ideas}. 
For example, P2-5 stated the difficulties of designing the scales of the dragon, ``\textit{we could not decide the suitable color for the scales}''. P2-4 noted that they ``\textit{did not know what materials to use for building a physical demo of the dragon'}'. 
The GenAIs can provide personalized guidance to address these challenges of idea implementation. 
For example, in Phase 2, the TA observed that students in Group 2 used GPT to figure out ``\textit{how to reinforce the body of the handcraft}''. P2-2 commented that the GenAIs ``\textit{can provide the most
suitable solution based on our actual situations}''.
\penguin{Additionally, three respondents mentioned that GPT was helpful in providing background knowledge and context about their projects.} 
For instance, P2-6 noted that GPT helped understand the ``\textit{characteristics of the dragons}''.

% \textbf{\penguin{Finding 2}: Students used GPT \penguin{for information seeking} and DALL-E \penguin{for visual inspiration} in a sequential order}. 
\textbf{\ReviseZQW{Finding 2: Students use GPT for information seeking and DALL-E for visual inspiration sequentially.}}
In Phase 2, we introduced GPT to the students. % for the first time. 
By analyzing the interaction logs and the accompanied of screen recordings, we found that out of the total 28 queries to GPT and DALL-E, 13 queries were posed in a two-step process. 
In the first step, students queried GPT for \penguin{information seeking}. 
% The types of these queries to GPT include practical guidance, 
For example, Group 1 asked, ``We want to make [the dragon named] Fu Xi. How can we use newspaper to make its head? We find this part very difficult''. 
Group 3 asked for information about the theme of the project, ``What does Bi Xi look like? He is one of the Nine Sons of the Dragon''.
Based on the information gathered from GPT, students then queried DALL-E for \penguin{visual inspiration} in the second step. 
For instance, Group 1 copied the complete answer from GPT and pasted it into the chat with DALL-E. 
Similarly, Group 3 asked DALL-E, ``Images of Bi Xi from the Nine Sons of the Dragon'' after getting a textual description of it from GPT.

% \penguin{
% \textbf{Finding 3: Students found it hard to type and formulate queries to GenAIs.} 
% } 
\textbf{\ReviseZQW{Finding 3: Students struggled with querying, from typing issues to uncertainty in what to ask, leading to unsatisfying GenAI responses.}} Based on the TAs' observations, students frequently face difficulties in typing down their queries to GPT and DALL-E, \penguin{which is a default input approach in Chatbox in Phase 2}. 
The TAs in Groups 1 and 2 recalled that they were requested to help with typing over 90\% of the time.
\penguin{Meanwhile, students were unsure of what questions were appropriate to ask to GPT and how to refine the queries}, as P2-4 said, ``\textit{I did not know what questions I could ask [to GPT] at that moment. I also did not know I could follow up; I thought that was it}''. 
\penguin{Consequently, students expressed dissatisfaction with the results returned by GenAIs, as indicated by three respondents in Phase 2. }
This dissatisfaction was further evidenced by \penguin{9 and 11} queries generating responses that were ``perceived as incorrect'' and ``perceived as correct but unhelpful''. 
For example, Group 1 asked for images about handcrafting with questions (\eg ``How to make the back of [the dragon] Fu Xi from the Nine Sons of the Dragon?'') in two out of their five queries to DALL-E, both of which were perceived incorrect by students. 
Group 2 asked GPT rather than DALL-E for ``the face of [the dragon] Qiu Niu''. 
Group 2 also inquired GPT about ``the color of [the dragon] Qiu Niu's pupils'' and, upon receiving an unsatisfactory answer, terminated the conversation and switched to using DALL-E.
Group 3 requested ``an image of [the dragon] Bi An'' from DALL-E without providing any description \peng{and expected to get background information of Bi An, as recalled by the TA in Group 3}.
Similarly, they asked GPT to provide ``an image of [the dragon] Suan Ni's handcraft'', which misinterpret GPT's abilities as an image generator. 

% \penguin{
% \textbf{Finding 4:} 
% }
% \textbf{Students desire step-by-step visual guidance in handcraft projects.} 
% During phase 2, students underscored the absence of step-by-step visual guidance, such as video tutorials, in GPT and DALL-E for executing their tasks. 
% P2-1 stated, ``although it [GPT] described detailed steps [for building a handcraft dragon], we wanted to see pictures showing each step''. This feedback highlights students' preference for visual aids that can simplify complex instructions and enhance understanding. Interestingly, students in Group 1 requested to use the TA's smart phone to search for a tutorial video on making the handcraft for Suanni (one of the Nine Sons of the Dragon) in Little Red Book, a popular platform similar to Instagram. They remarked, “it [GPT and DALL-E] can not provide us with videos, while Little Red Book can”. This positions promising future opportunities for assisting elementary school students’ art projects with generated videos and / or user-generated content (e.g., how-to videos) in social media platforms.

\penguin{
\section{\name{}: Customizing Interactions with GenAIs for Facilitating Elementary Students in Art Projects}
\label{sec:askart_design}
The findings in Phases 1 and 2 revealed the benefits of GenAIs for facilitating elementary students' information seeking and solution ideation in group-based art projects (Finding 1). 
However, students reported challenges in typing and formulating queries to GenAIs (Finding 3). 
Together with the students' patterns of  interacting with GPT and DALL-E (Finding 2), our research team compiled three design goals and came up with interaction features with GenAIs for supporting students in PBL art courses. 
Our team consists of four master students majored in human-computer interaction (HCI) and one assistant professor specialized in HCI, and for one month, they met weekly to iteratively discuss, refine, and implement these interaction features in an interface we called \name{}. 
The three design goals of \name{} are as below. 
}
% \subsection{Design and Implementation of \name{}: A Customized GenAI Interface for Facilitating K-6 Students in Art Projects} \label{sec:askart}

% \pzh{
% Our analyses of the data collected in Phases 1 and 2 reveal several students' challenges and interaction patterns in using DALL-E and GPT in their art projects. 
% In the next section Findings (\autoref{sec:findings}), we report these results together with those from Phases 3 and 4, as many challenges and patterns consistently appear across the four phases. 
% As reported later in \autoref{sec:findings}, some of the challenges (\eg hard to type and formulate queries to GenAI (\autoref{sec:challenges})) and interaction patterns (\eg use GPT and DALL-E in sequential order (\autoref{sec:challenges})) in Phases 1 and 2 informed our design of \name{}, a customized GenAI interface for facilitating K-6 students in their group art projects. 
% In this subsection, we do not repeat these results but describe the design guidelines, interface, and implementation of \name{}.

% \subsubsection{Design Guidelines}

% Based on the analysis of the challenges and user patterns observed during the study, several 
% The design of \name{} is guided by three principles after observing students' challenges and patterns of using DALL-E and GPT in Chatbox for their art projects. 
% design guidelines were identified to enhance the usability and effectiveness of GenAI tools in K-6 project-based learning (PBL) art courses:

\textbf{DG1: \name{} should simplify students' input and tailor the initial prompts to GenAI based on the theme of the projects}. 
% The interface should be designed to be user-friendly for K-6 students, featuring simplified input methods and prompt organization tailored for young learners. 
This could reduce the complexity of typing prompts for elementary school students and improve the relevance of generated content to the art projects. 

\textbf{DG2: \name{} should support a seamless transition from the content {\ReviseFinal{{\ReviseFinal{generated by a text-based AI model to that produced by}}}} the image generator}. 
% Ensure seamless integration of GPT and DALL-E to support a smooth workflow that aligns with user patterns. 
This involves using GPT for text-based information gathering and then utilizing DALL-E for visual creation based on the gathered information. This transition between the usage of GPT and DALL-E should be intuitive and smooth. 

\textbf{DG3: \name{} should provide clear and effective guidance for prompting the GenAI models.}
Clear guidance can help users correctly understand and use GPT and DALL-E. This includes example use cases, templates, and adaptively suggested prompts that highlight the unique capabilities and best practices for each model.

\pzh{
\subsection{\penguin{\name{} Design and Implementation}}

\begin{figure}[h]
  \centering
  \includegraphics[width=\linewidth]{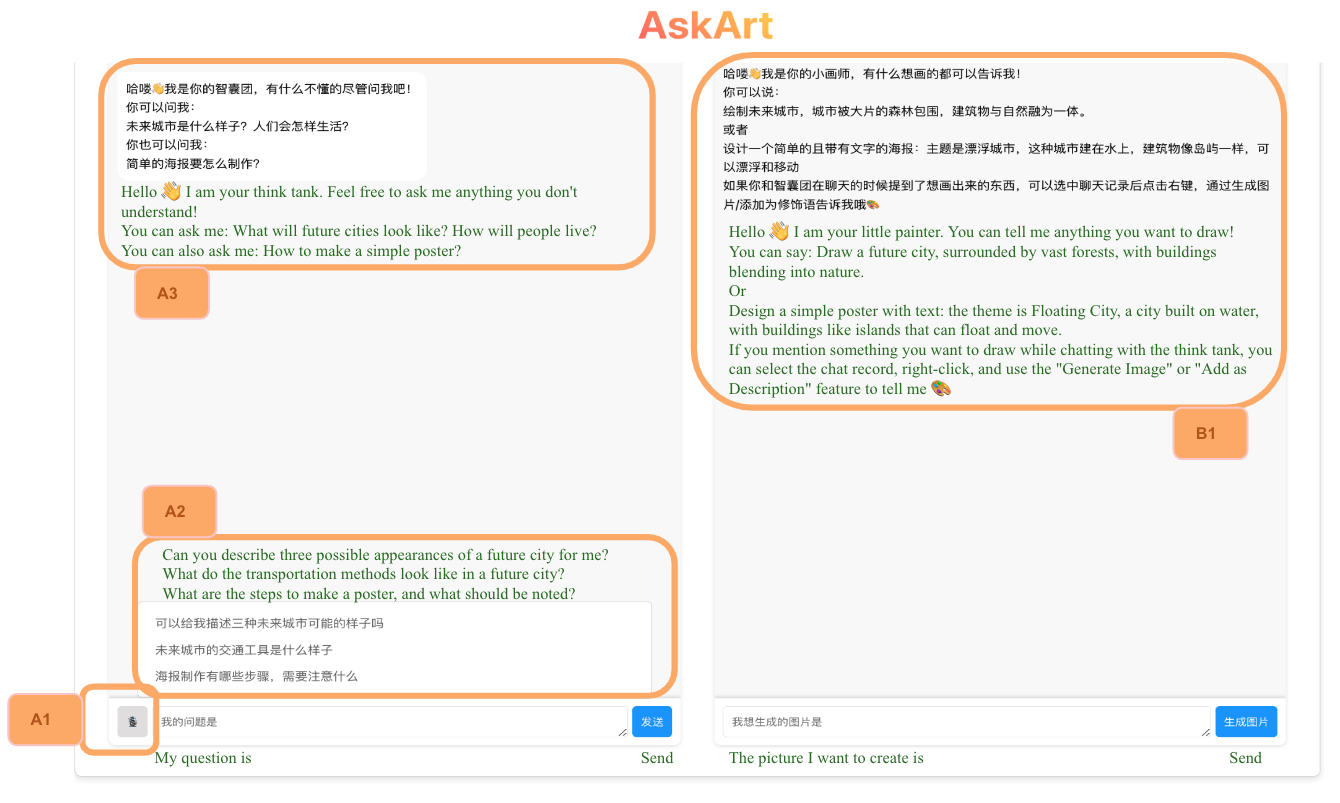}
  \caption{\name{} Interface for Coldstart. The left part is the chat interface with ``\textit{Think Tank}`` (powered by GPT), featuring: Audio-to-Text input (A1), Suggested Follow-Ups (A2), and Project-Related Introduction of ``\textit{Think Tank}`` (A3). The right part is the chat interface with ``\textit{Little Painter} ``(powered by DALL-E), featuring: Project-Related Introduction of ``\textit{Little Painter}`` (B1).}
  \label{fig:askart1}
  \Description{}
\end{figure}

\begin{figure}[h]
  \centering
  \includegraphics[width=\linewidth]{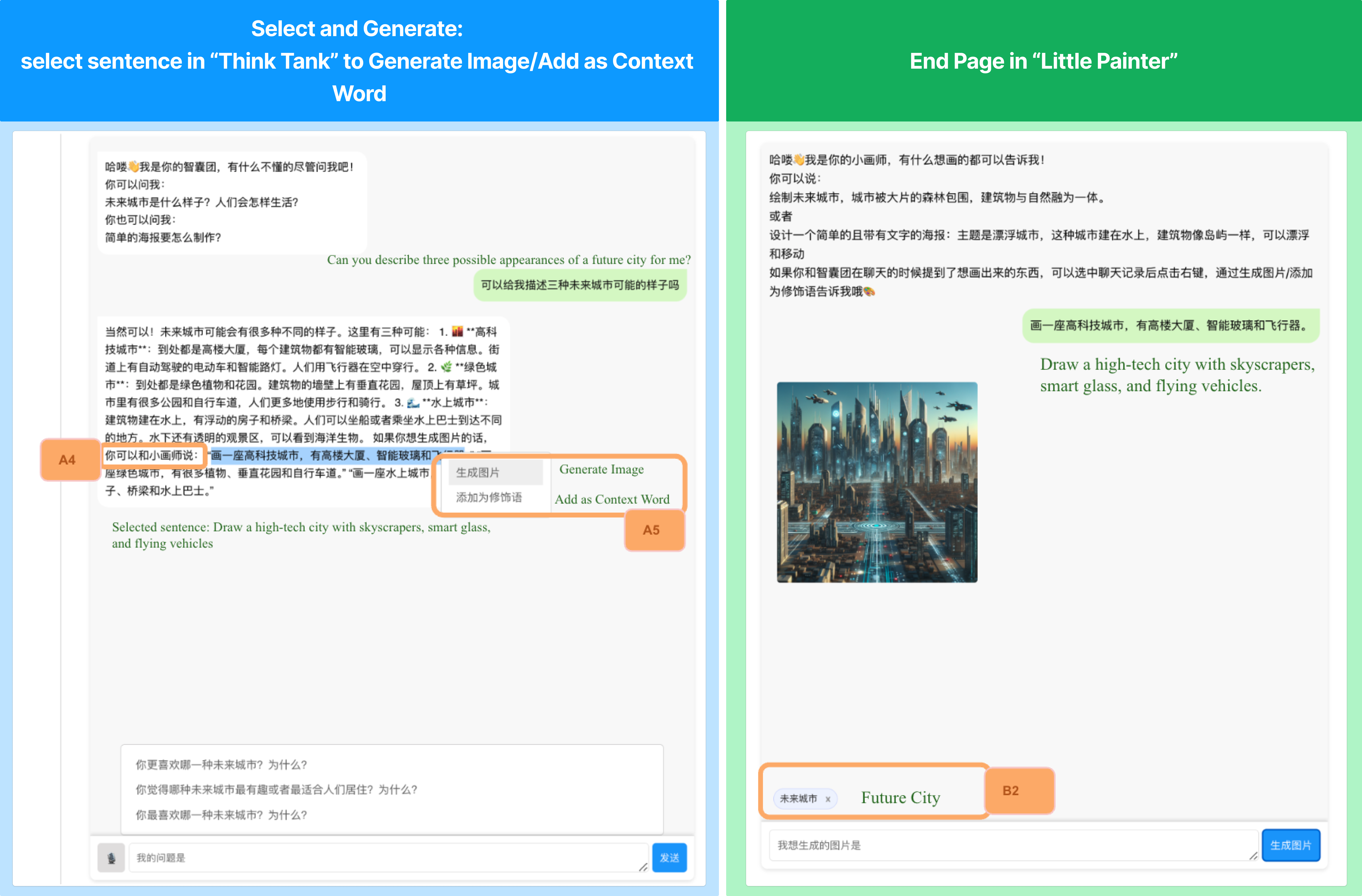}
\caption{\name{} Interface for \textit{Auto-Generated DALL-E Prompts} and \textit{Select and Generate}. The blue section is the interface of ``\textit{Think Tank}``, showing the selection of \textit{Auto-Generated DALL-E Prompts} (A4) followed by a right-click to use \textit{Select and Generate} (A5). The green section is the end page in the interface of ``\textit{Little Painter}``, demonstrating the resulting image sent by ``\textit{Little Painter}`` after the actions in the blue section. B2 represents the \textit{Context Word}.}
   \label{fig:askart2}
  \Description{}
\end{figure}

% The design goals for AskArt were translated into a comprehensive feature framework aiming at enhancing the usability and effectiveness of GenAI tools for K-6 students in project-based learning (PBL) art classes.
\pzh{\autoref{fig:askart1} and \autoref{fig:askart2} show the screenshots of the \name{}'s interface}. The interface is divided into two main sections: ``\textit{Think Tank}'' powered by GPT for text generation \pzh{(\autoref{fig:askart1} left section)}, and ``\textit{little Painter}'' powered by DALL-E for text-to-image generation \pzh{(\autoref{fig:askart1} right section)}. 
In each section, users can scroll the chat window up and down. %This layout supports independent scrolling for each section, facilitating smooth navigation and usage.
\name{} has the following unique features upon Chatbox based on the design goals.

\textbf{Project-Related Introduction of GenAI}. At the cold start, both ``\textit{Think Tank}'' and ``\textit{Little Painter}'' will send out introductory messages(\autoref{fig:askart1} A3 B1). 
These messages introduce the basic functionalities and user cases tailored to the PBL theme and tasks (\textbf{DG3}).  
The ``\textit{Think Tank}'' interface also features two initial prompts in the Follow-Up Questions section, designed to help students understand its role and provide starting points for interaction \pzh{(\autoref{fig:askart1} A2)}. 
These introductions and initial prompts were carefully crafted by our research team and teachers, and they were manually inputted before Phases 3 and 4 based on the corresponding project themes and tasks. %  \shiwei{(shown in Table xx in Appendix)} 
% We ensured that the designs took into account the different themes and tasks of the PBL, addressing the potential variety of questions that could arise.

\textbf{Audio-to-Text Input}. 
% To address the difficulty students face with typing \textbf{(C1)}, 
We implemented an audio-to-text input feature in the ``\textit{Think Tank}'' section \pzh{(\autoref{fig:askart1} A1)}. 
This feature allows students to vocalize their queries, which are then converted to text \peng{using Web Speech API}. %, easing the typing burden and facilitating faster interaction with the AI tools. 
This feature could ease the students' input effort to GPT \textbf{(DG1)}.
% \peng{We do not }

\textbf{Suggested Follow-Ups}. To encourage students to seek further information or clarification, we introduced ``\textit{Suggested Follow-Ups}'' at the top of the GPT input bar \textbf{(DG3)}. % \textbf{(C5)}. 
These suggestions are generated based on the current conversation and update dynamically as the dialogue progresses. 
This feature aims to make students aware that they can provide feedback or refine their requests, and it also brings new angles for inquiry that students may not have considered, which could foster creativity and thoroughness. 
% This aligns with the guideline for providing clear guidance and usage tips \textbf{(DG3)}.

% \textbf{Auto-Generated DALL-E Prompts}.
% To provide a smooth transition between the usage of GPT and DALL-E \textbf{(DG2)}, we introduce a feature where each GPT response in ``\textit{Think Tank}'' interface ends with an automatically generated draft prompt for DALL-E in ``\textit{Little Painter}'' interface \pzh{(\autoref{fig:askart1} A4)}. This prompt is based on the content of the corresponding GPT response, allowing students to directly use it if they wish to visualize the information provided by GPT. 
% This not only aligns with the user pattern of sequentially using GPT and DALL-E but also dynamically demonstrates to students how to construct effective queries for DALL-E. 
% Offering a ready-to-use prompt could also simplify the input process (\textbf{DG1}) and encourage students to visualize their text-based findings immediately \pzh{(\autoref{fig:askart2} A4)}. %, thereby enhancing their learning experience (see Fig. 2). 
% This feature addresses the challenge of missing context when interacting with DALL-E \textbf{(C3)} and adheres to the guidelines for seamless tool integration \textbf{(DG2)} and user-friendly interfaces \textbf{(DG1)}.

% \textbf{Select and Add Content in GPT as the Prompts to DALL-E}. 
\penguin{\textbf{Select and Generate}}. 
Another key feature in \name{} is the ability for students to select sentences or words from the ``\textit{Think Tank}'' chat to use as prompts or parts of the prompt for ``\textit{Little Painter}'' (\textbf{DG2}). %, named as ``\textit{Select and Generate}'' (\textbf{DG2}). 
\penguin{
To help students use this feature, each GPT response in ``\textit{Think Tank}'' interface ends with an automatically generated draft prompt for DALL-E in ``\textit{Little Painter}'' interface \pzh{(\autoref{fig:askart1} A4)}.} 
This prompt is based on the content of the corresponding GPT response, allowing students to directly use it if they wish to visualize the information provided by GPT. 
This not only aligns with the user pattern of sequentially using GPT and DALL-E \penguin{(Finding 2)} but also dynamically demonstrates to students how to construct effective queries for DALL-E \penguin{(Finding 3)}. 
Offering a ready-to-use prompt could also simplify the input process (\textbf{DG1}) and encourage students to visualize their text-based findings immediately \pzh{(\autoref{fig:askart2} A4)}.
After selecting the desired text, students can right-click to access a menu \shiwei{(\autoref{fig:askart2} A5)} with two options:
\begin{itemize}
\item \textit{Generate Image}: Selected content is copied directly into DALL-E's input box, indicating the main subject for visualization.

\item \textit{Add as Context Word}: Selected content appears above DALL-E's input box and can be manually deleted if needed. 
This ensures that essential thematic or stylistic elements are included in the visualization, preventing omissions that might lead to unsatisfactory results. 
Context words can include PBL theme information or specific elements \shiwei{about art},  
such as colors and styles that students want to incorporate into their artwork. 

\end{itemize}
This structured approach can prevent students from overloading the input box in DALL-E with all information, clearly distinguishing between the main subject of the image and additional contextual details. 
This feature can also encourage students to include relevant background information and stylistic preferences, which would improve the accuracy of DALL-E's visual output (\textbf{DG3}). 
% These features address the challenges of missing context \textbf{(C2)} and wasted GPT information when interacting with DALL-E \textbf{(C3)}, and align with the guidelines for user-friendly interfaces \textbf{(DG1)}, coherent tool integration \textbf{(DG2)}, and guided usage \textbf{(DG3)}.
}

\pzh{
% \subsubsection{Implementation of \name{}}
\name{} is developed using the Vue.js framework for the front end, and Python with Flask for the back end. 
Data is stored locally on the website for improved performance and quick access. 
The system processes user inquiries related to art by sending requests from the Vue.js interface to the Flask server, which then interacts with OpenAI's GPT-4o and DALL-E 3 API to generate responses. These responses are processed and returned to the user through the Vue.js front end. The project ensures that all user interactions and queries are recorded and utilizes real-time data to provide accurate information.
}

\begin{figure}[h]
  \centering
  \includegraphics[width=\linewidth]{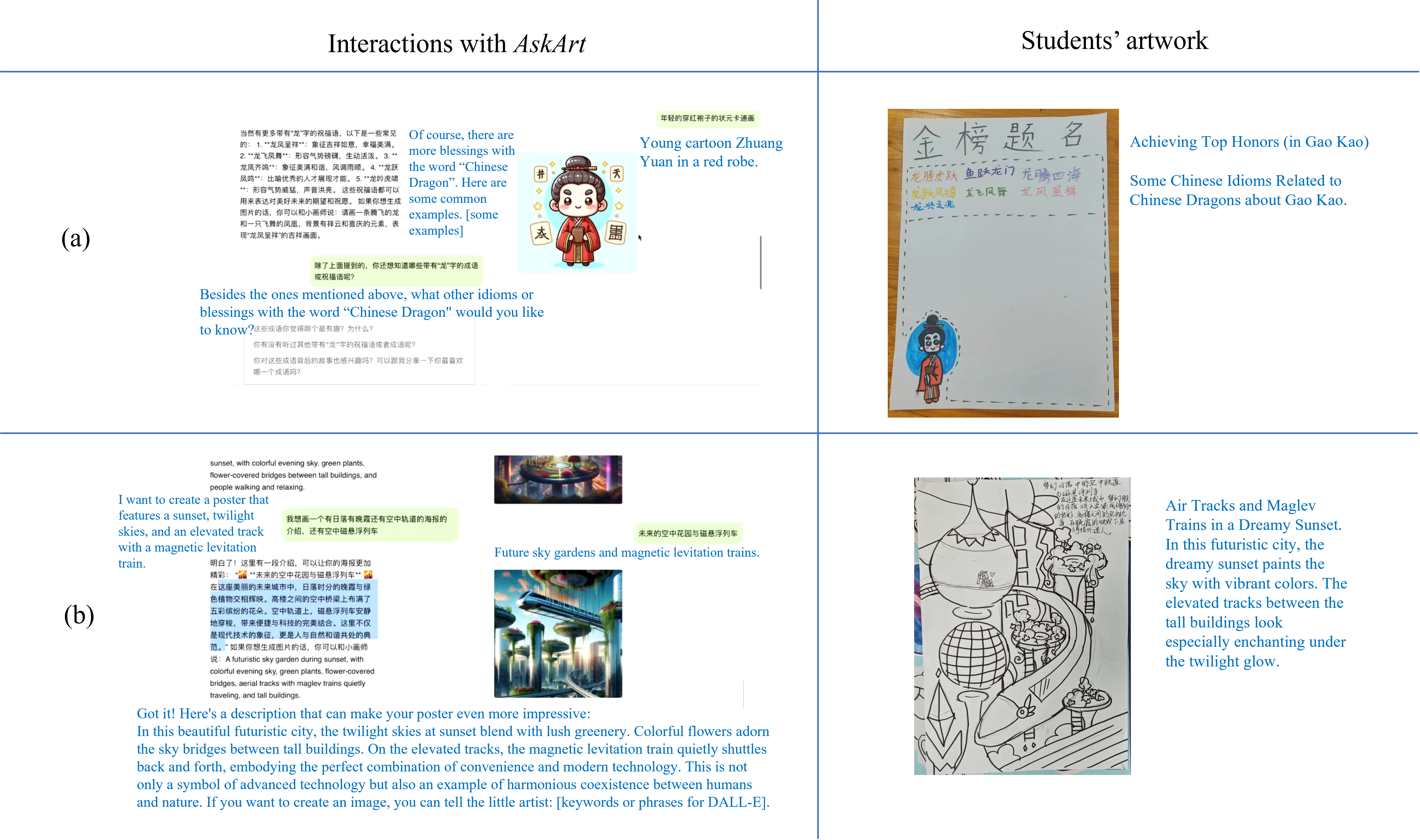}
\caption{The samples of students' interactions with \name{} and the corresponding outcome. (a) Group 2 in Phase 3. (b) Group 1 in Phase 4.}
     \label{fig:sample_artwork}
  \Description{}
\end{figure}

\penguin{
\section{Phases 3 and 4: Evaluation of \name{}}
\label{sec:phase_3_4}
After developing \name{}, we collaborated with Teacher 1's teaching team and evaluated \name{} in their offline art courses. 

\subsection{Procedure, Project Task, and Collected Data}
\textbf{Phase 3} happened in Week 14 of 2024 Spring semester. 
In the interview with Teacher 1 before the course session, we introduced \name{} to the teacher and gathered her initial feedback. 
% Teacher 1 suggested that we could allow students to use \name{} in her intended course session for 
Different from Phases 1 and 2, as the semester progressed, the task for each student group changed to ``using dragon-related text and images to create a poster that blesses the high-school students in `Gao Kao', the college entrance exam in China'' (Fig. \ref{fig:sample_artwork}(a)). 
Meanwhile, similar to Phases 1 and 2, this task requires students to work in group to ideate and implement an artwork (poster, specifically) as the outcome, during which we can examine how GenAIs would help. 
In the course session, 
}
similar to Phase 2, three members of the research team acted as TAs in three student groups, each with a laptop that have access to \name{}. 
At the end of the course session, we distributed Questionnaire 3 to six randomly selected students, two from each group, and collected five responses, as one student was not available to complete the questionnaire at that time. 
Questionnaire 3 asks: i) challenges encountered during class; ii) benefits and challenges of using \name{}; iii) ratings for GenAIs' assistance in information seeking and implementation guidance (1 - strongly disagree, 7 - strongly agree); iv) ratings for the assistance provided by \name{}'s specific features: \textit{Audio-to-Text Input}, \textit{Suggested Follow-Ups}, \textit{Select and Generate} (1 - strongly disagree, 7 - strongly agree).
We conducted quick interviews with the five respondents to make sense of their responses to the questionnaires. 
We also interviewed Teacher 1 after the course session. We discussed students' perceptions of GPT, DALL-E, and \name{}, assessed the impact of GenAI in these sessions so far, and discussed potential ways to improve \name{}. 
The \textbf{data collected in Phase 3} include: audio recordings of the interviews with Teacher 1 and students; 
Students' prompts to GPT and DALL-E;  
responses of five students on Questionnaire 3;
and screen recordings of \name{}'s usage in the computers (with the audio input and camera on) for the three student groups. 

% \subsection{Phase 4: Students' Perceptions in \name{} Groups and Baseline Groups without GenAI} \label{sec:phase4}
\penguin{Phase 3 could help us qualitatively examine the specific features of \name{} against Chatbox, as the students in Class A used both tools in group-based projects with a similar solution ideation process. 
However, it can not enable us to evaluate \name{}'s impact on the students' PBL process compared to the cases without \name{}. 
\textbf{Phase 4} sought to address this limitation with a between-subjects study design. 
}
% In Phase 4, we work with Teacher 2 and a new set of students with a refined experimental setup. 
% The participants included the another elementary school art teacher(Teacher 2) from earlier phases, who has 20 years of art education experience and 8 years of experience applying project-based learning (PBL) in art education. 
Specifically, different from those in Phases 1 to 3, students in Phase 4 were all from grade 6 and were assigned to Class B and Class C, both with 54 active students (29 males and 25 females).
Each class was divided into 8 groups, \pzh{each with 6-8 students}. 
\penguin{We collaborated with Teacher 2, who was responsible for the art courses of Classes B and C, to set up our study. }
In the interview with Teacher 2, before the course session, the research team introduced \name{} to the teacher and gathered initial feedback. 
They decided to conducted the study in week 16 of 2024 Spring semester, in which the course theme is ``\textit{Future City Poster}''. 
% \penguin{In this course session, the task for each student group was to create posters depicting a future city}. 
% Again, three members of the research team acted as TAs in three student groups in each class, each with a laptop that has access to \name{}. 
% \peng{Different from Phases 2 and 3, in Class B or Class C, the three groups using \name{} were randomly selected, and the other five groups can not access \name{} nor other computer devices}. 
\ReviseZQW{
In this course session, the task for each student group was to create posters depicting a future city (Fig. \ref{fig:sample_artwork}(b)). A total of eight groups were formed in each class (Class B and Class C), with six groups assigned to the experimental condition and ten groups to the baseline condition. For the experimental condition, each group used AskArt. For the baseline condition, no additional tools were provided, reflecting the standard classroom setting for normal PBL classes without GenAIs, as advised by the teachers. Teachers typically allow students to access tools like search engines at home but avoid providing such tools during class sessions due to concerns that access to search engines and social media platforms might distract students with unrelated content and hinder their focus.}

\Original{
At the end of the course session in Phase 4, we distributed Questionnaire 4-all to all students and collect \peng{108} responses. 
\textbf{Questionnaire 4-all} asks for ratings on i) overall challenge in the class; ii) overall satisfaction; iii) participation in team collaboration; and iv) expression of creativity in class; 1 - the lowest score, 7 - the highest score. 
We also distributed Questionnaire 4 to the students in groups with \name{} and collect \peng{13} responses. 
\textbf{Questionnaire 4} uses the same questions as Questionnaire 3, except for i) challenges encountered during class, to avoid repetition with Questionnaire 4-all. 
\peng{After the course session, we conducted quick interviews with in total of 3 respondents to make sense of their responses to the questionnaires}. 
Lastly, we interviewed Teacher 2, discussing students' perceptions of GPT, DALL-E, and \name{}, assessing the impact of GenAI in this course session, and discussing potential ways to improve \name{}.
The \textbf{data collected in Phase 4} include: audio recordings of the interviews with Teacher 2 and students;
students' prompts to GPT and DALL-E;
responses from all 108 students on Questionnaire 4-all and responses from 13 students from the groups with \name{} on Questionnaire 4;
and screen recordings of \name{}'s usage in the computers (with the audio input and camera on) for the three student groups. }

\penguin{
\subsection{Analysis and Results of Phases 3 and 4}
}
% \penguin{[To be continued]}
\penguin{
\subsubsection{Data Analysis} 
Similar to \autoref{sec:data_analysis}, we employed the reflex thematic analysis approach to analyze the audio recordings of interviews with teachers and students collected in \penguin{Phases 3 and 4}. 
The outstanding themes from the interview with teachers answer the \textbf{RQ3}, while those from the interview with students complement the Findings 1-3 of \textbf{RQ1} and answer \textbf{RQ2}. 

For the questionnaire data, we first filtered unsuitable responses by carefully reviewing each response. }
For Questionnaire 4-all, we observed that some students left comments under the rating questions. If there was a mismatch between the attitudes revealed in the comment and the rating score (\eg, a negative comment for a question rated 7 out of 7), these responses were filtered out, resulting in the exclusion of 9 out of 108 questionnaire sheets.
For the students' ratings on the remained Questionnaire 4-all, we performed unpaired t-tests to statistically compare the differences between the groups (43 valid respondents) with and without (54 valid respondents) \name{} regarding students' perceived challenges, satisfaction, team collaboration, and creativity in the course session.

For the prompts to GPT and DALL-E and screen recordings, 
\penguin{we used the similar analysis approach in \autoref{sec:data_analysis} to code the user behaviors and interactions with \name{}.} 
Students' responses to the answers generated by GenAIs in \name{} were also classified into three categories: (1) perceived as correct and helpful, (2) perceived as correct but not helpful, and (3) perceived as incorrect. 

\begin{figure}[h]
  \centering
  \includegraphics[width=\linewidth]{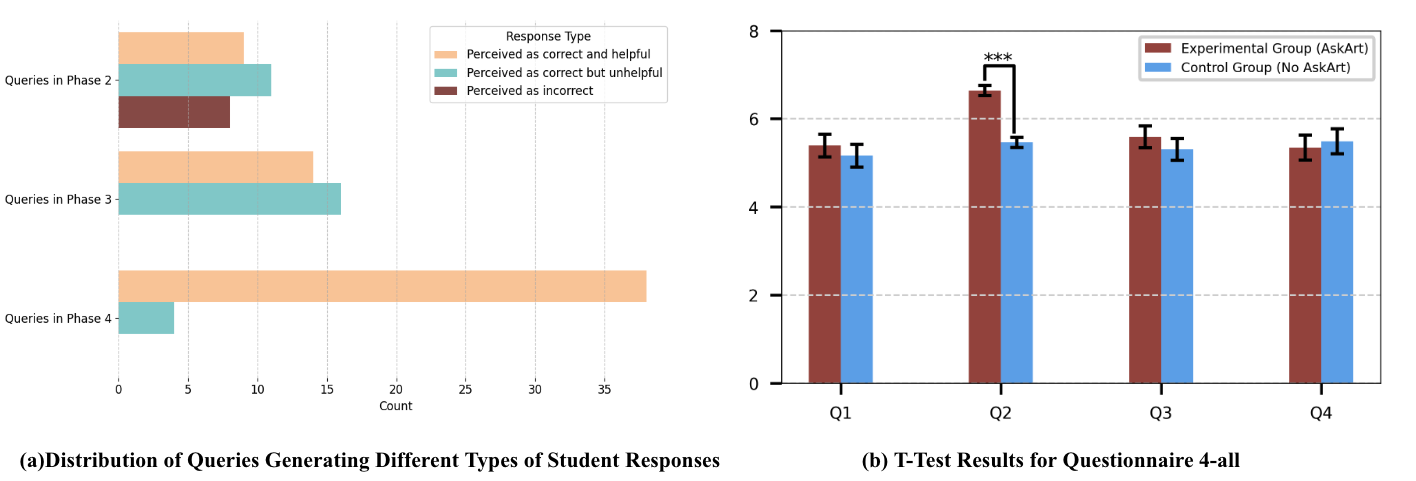}
\caption{Results for Content Analysis for Students' Prompts in Phases 2-4 and t-test in Phase 4. (a) shows the responses of students to the answers generated by GenAI based on their queries, with bars representing the number of queries ending in three types of responses. (b) shows the t-Test results comparing  n student ratings of their PBL class experiences between those who used AskArt (experimental group) and those who did not (control group). \ReviseZQW{Q1 is ``How challenging did you find this lesson?'' (1: very low, 7: very high); Q2 is ``Rate your satisfaction with this class.''(1: very unsatisfied, 7: highly satisfied); Q3 is `` actively participated in team collaboration during this lesson.''(1: strongly disagree, 7: strongly agree); Q4 is ``I fully expressed my creativity in this less. ''(1: strongly disagree, 7: strongly agree).} The results indicate a significant difference in overall satisfaction (Q2) between the groups, with the experimental group rating higher. Error bars represent standard deviations, with *** indicating a p-value < 0.001.}
     \label{fig:findings1}
  \Description{}
\end{figure}

\penguin{
\subsubsection{
Findings in Phases 3 and 4 for RQ1
}
Students' responses to the questionnaires in Phases 3 and 4 complemented \textbf{Finding 1} (\autoref{sec:finding_phase_1_2}) about GenAIs' helpfulness in inspiring project ideas, guiding idea implementation, and providing background information about the project. 
For example, in Phase 4, 
}
P4-13 commented the benefits from GenAIs as ``it provides us with creative ideas for the project''. 
The teaching assistant's (TA) observation of Group 3, where P4-13 belonged, along with this group's usage logs of \name{}, illustrates how GenAIs inspired their project. During the brainstorming session, they initially inquired with GPT about what a ``Future City'' might look like. They were particularly inspired by several elements mentioned by GPT, especially the idea of ``hanging gardens''. Subsequently, they used DALL-E to generate a visualization of a ``Future City with hanging gardens'' through the \textit{Select and Generate} feature. The generated image further inspired them, leading to the inclusion of ``hanging gardens'' in their final poster. 
% \penguin{Besides, students noted the values of GPT in seeking background information.} 
Besides, P3-2 in Phase 3 emphasized the effectiveness of GPT in providing background information about the theme of the project, particularly when he ``\textit{searched for blessings related to dragons}''. 
P3-4 noted that he/she learned from GPT's answers about ``\textit{idioms related to dragons}''.
In Phase 4, P4-2 highlighted that the GPT enabled the group to ``\textit{quickly understand the project's theme about future city}''.

\ReviseZQW{In Questionnaire 4-all (Fig. \ref{fig:findings1}), students using AskArt ($M = 6.64, SD = 0.12$) reported significantly higher ($t = 3.62, p = 0.0005$) overall satisfaction compared to those in the baseline group ($M = 5.46, SD = 0.27$). However, there was no significant difference ($t = 0.57, p = 0.57$) in perceived challenge level between the experimental group ($M = 5.39, SD = 0.26$) and the baseline group ($M = 5.16, SD = 0.28$). Similarly, for creativity expression, there was no significant difference ($t = -0.37, p = 0.72$), and the baseline group reported a slightly higher mean score ($M = 5.49, SD = 0.28$) compared to the experimental group ($M = 5.34, SD = 0.28$).
}

\penguin{
In Phase 4, students with \name{} in Classes B and C did not use GenAIs in their offline art courses before. 
We observed their interaction patterns with \name{} at a group level and got \textbf{Finding 4: Student groups either assigned fixed members or allow every member to interact with \name{} in the art courses}. 
}
For example, in Group 2, the emerging group leader assigned two students to interact with \name{}. These students were responsible for collecting questions from the rest of the group, formulating and posing queries to \name{}, and presenting the generated answers back to the group. Using this strategy, Group 2 still left some students, who were not engaged in group collaboration, uninvolved in the interaction with \name{}.
In contrast, in Groups 6, any student with a question can interact with \name{}. This back-and-forth interaction with GenAIs allowed for individual adjustments based on the AI's suggestions. If the suggested adjustments impacted the whole design, the student would show the related GenAIs answer and discuss within the group; otherwise, they would make the decision independently and keep the GenAIs answer to themselves without sharing to the group. In this case, students, who were not engaged in group collaboration or did not pose any queries to \name{}, were uninvolved in the interaction with \name{}. \ReviseZQW{Team engagement ratings (Fig. \ref{fig:findings1}(b)) further support these observations, showing no significant difference ($t = 0.70, p = 0.48$) between the experimental group ($M = 5.59, SD = 0.24$) and the baseline group ($M = 5.31, SD = 0.30$).}

Findings 2 and 3 (\autoref{sec:finding_phase_1_2}) in Phases 1 and 2 informed the specific features of \name{}, and we found that these features streamlined the students' interactions with GPT and DALL-E. 
We described findings related to each feature in the next subsubsection. 

\penguin{
\subsubsection{
Findings for RQ2
}
Our evaluation studies of \name{} provided empirical understandings to \textbf{RQ2} (Fig. \ref{fig:summary_of_findings}), \ie ``What interaction design with GenAIs would be helpful for supporting students in PBL art courses?''
}

\penguin{
\textbf{Finding 5: The ``Project-Related Self-Introduction of GenAIs'' improved students' understandings and proficiency of using GenAIs}. 
}
\autoref{fig:findings1} (a) summarizes our analyses of the students' interaction logs with GenAIs. 
Specifically, we calculated the number of students' incorrect usages of GPT and DALL-E, represented by the number of queries generating ``perceived as incorrect'' responses (e.g., using DALL-E for instructional steps and GPT for image information). 
Our findings suggested that providing students with an introduction and user cases of GenAIs at the interface could improve their ability to use GPT and DALL-E correctly. This improvement is indicated by the disappearance of queries generating ``perceived as incorrect'' responses in Phases 3 and 4 (\autoref{fig:findings1} (a)). 
The introduction described GPT as a ``think tank'' and DALL-E as a ``little artist'', which helped students distinguish that GPT was not meant for visualization and DALL-E was not for textual information seeking. 
No instances of incorrect usage for GPT and DALL-E were observed in both Phase 3 in which students with prior experience using GenAIs in previous phases and Phase 4 in which students used GenAIs for the first time. 
For example, in Phase 3, students in Group 1 queried GPT for textual answers, \eg ``Are there any idioms related to dragons?'', and prompted DALL-E to visualize their ideas, \eg ``A cool version of Ying Long'', which demonstrates their improved understanding of the GenAIs' capabilities.

\penguin{
\textbf{Finding 6: The ``Audio-to-Text Input'' to GPT was disorganized or incomplete, which was not helpful for supporting students in PBL art courses.}
}
In Phases 3 and 4, we introduced \name{} that allows students to query the GPT via voice input (\autoref{fig:askart1} A1). 
Students actively used this feature in 16 out of a total of 25 queries to GPT. However, \peng{we observed} that in many cases, the voice input was disorganized or incomplete. 
For example, in Phase 3, students in Group 3 spoke ``blessings related to dragons'' to \name{}, \peng{which output many positive idioms that contain the Chinese word (\ie ``Long'') about dragon for New Year's (\eg Good luck in the Year of the Dragon -- in Chinese: Long Nian Da Ji), birthday, and wedding blessings (\eg prosperity brought by the dragon and the phoenix -- in Chinese: Long Feng Cheng Xiang).}
However, what these students needed were the blessings related to dragons for passing important exams.
A complete query to GPT for this purpose would be ``we are making a poster about passing important exams and need blessings related to dragons'', which would output Chinese idioms like ``Li Yue Long Men'', wishing people, like the carp leaping over the dragon gate, achieve excellent results in the examinations and have a bright future.
Similarly, in Phase 4, students in Group 6 talked to \name{}, ``What words should be written in the painting of the Hanging Gardens?'', while the organized and completed prompt, as observed by the TA, should be ``I am making a poster of the Hanging Gardens. What slogan should be written on it?''. 
% Feedback from TAs during Phases 3 and 4 supported these observations, indicating that students' voice input frequently required further organization and clarification. 
One TA highlighted that this issue was particularly common when students were either nervous using the audio input feature for the first time or posed questions without organizing their thoughts beforehand. 
The TA explained ``\textit{Sometimes I needed to reconfirm the intended meanings of their voice input, which, of course, can not be understood by GPT}''. 
These observations emphasize the ongoing need for improving the voice input to GenAIs (\eg via auto-completion of the transcribed text) to make them easier to use for elementary school students, \peng{which we will discuss in the Discussion section}.

\penguin{
\textbf{Finding 7: The ``Suggested Follow-Ups'' helped elementary students engage with GPT for seeking needed information.} 
}
\name{} offers a follow-up question panel (\autoref{fig:askart1} A2) above the input bar. 
\peng{The respondents of Questionnaire 3 and Questionnaire 4 highly appreciated the helpfulness of this panel, with an average score of 6.06 ($SD = 1.51$); 1 - not helpful at all, 7 - very helpful}. 
For example, in Phase 3, Group 2 utilized the follow-up questions in the ``Suggested Follow-Ups'' feature twice. First, they asked for more blessings containing the word ``dragon'', and then they requested additional examples that did not repeat the previous answers. This sequence of follow-ups concluded with the students perceiving the final response as correct and helpful (\autoref{fig:findings1} (a)).
% \peng{??? The first selection resulted in an R2 rating (correct but not helpful), while the second selection met the students' expectations, earning an R1 rating (correct and helpful). }
Additionally, students began independently providing feedback. 
We also observe that students in Group 1 provided feedback to the GPT's responses. 
They initially queried GPT with ``Are there any idioms related to dragons for the college entrance exam?''. 
When finding the response inadequate, they further queried GPT in the next turn with ``I need more relevant examples''. 
\peng{GPT has the ability to conduct such multi-turn conversation and provided Group 1 with more related idioms.}

\penguin{
\textbf{Finding 8: The ``Select and Generate'' feature helped students iterate their queries to DALL-E.}
}
\name{} enables users to select words or sentences from the chat with GPT and use them as \textit{Context Word} for generating images in DALL-E (\autoref{fig:askart2}). 
Respondents of Questionnaire 3 and Questionnaire 4 rated this feature highly helpful ($M = 6.12, SD = 1.37$); 1 - not helpful at all, 7 - very helpful.
During Phase 3, students in Group 1 utilized this feature to make use of answers in GPT to form 5 out of 7 queries to DALL-E, and students in Group 3 utilized it for 5 out of 8 queries to DALL-E. An example of students using this feature in Phase 4 is shown in \autoref{fig:sample_artwork} (b). 
In Phases 3 and 4, we observed that students iteratively adjusted the main prompts and \textit{Context Word} (\autoref{fig:askart2} B) to DALL-E based on the images generated in the last turn. 
For example, in Phase 3, students in Group 1 iteratively refined their queries by making specific changes at each step. They began with ``dragon's roar and phoenix's cry'' without any \textit{Context Word}. 
They then added ``a handsome version of Yinglong'', incorporating ``dragon's roar and phoenix's cry'' as a \textit{Context Word}. 
Next, they changed it to ``a cute version of Ying Long with wings'', adding ``majestic dragon and phoenix's grace'' along with the existing ``dragon's roar and phoenix's cry'' as \textit{Context Word}. 
Finally, they settled on ``a handsome version of Ying Long with wings'', further adding ``cloud dragon and wind tiger'' to the list of \textit{Context Word}. 
Similarly, Group 2 students started with the query ``related to the [dragon] Zhuang Yuan''. They refined this by changing it to ``Zhuang Yuan in a red robe''. Next, they modified it to ``cartoon Zhuang Yuan in a red robe'', and finally, they specified it further to ``young cartoon Zhuang Yuan in a red robe'' (\autoref{fig:sample_artwork} (a)). 
This iterative process indicates that students can progressively enhance their queries by adding and adjusting \textit{Context Word} to get needed images from DALL-E.

\penguin{
\subsubsection{
Findings for RQ3
}\label{sec:rq3}
The interviews with Teacher 1 in Phases 1-3 and Teacher 2 in Phase 2 contributed understandings to RQ3 from the teachers' perspectives. }
\penguin{
We summarized the two teachers' perceived values, impact, and concerns of GenAIs on the students' PBL process after running the course sessions (Fig. \ref{fig:summary_of_findings}). 
}

\penguin{\textbf{Finding 9: Teachers welcomed the usage of GenAIs for information seeking and solution ideation in art projects}}. 
\penguin{First of all, before running the course sessions,}
both teachers expressed strong support for and interest in integrating GenAIs interfaces into the students' project-based learning (PBL) process. Teacher 1 highlighted the relevance of GenAIs to PBL and art courses, ``GenAIs are incredibly powerful and already widely used in professional contexts. It aligns well with the inquiry-based nature of PBL, and I am very optimistic about its integration into both group project-based art courses and other type of art courses in the future''. 
Teacher 1 also noted that when interacting with GenAIs, \penguin{especially in the \name{} interface}, students were less likely to be distracted by irrelevant information \penguin{compared to using web search engines for information seeking}. 
``\textit{In previous classes [without GenAIs], when we use computers to open web pages, we can not ensure that the pages only show information related to our courses, and students are easily distracted. GenAIs helped us mitigate this issue in the recent courses}''.

\penguin{\textbf{Finding 10: DALL-E provided a good starting point for students with deficiencies in drawing or a lack of imagination, and GPT improved students' efficiency in completing projects.}} 
Teachers mentioned that with the assistance of GenAIs, students demonstrated an increasing engagement in completing the challenging projects. 
For students with deficiencies in drawing or a lack of imagination, DALL-E provided a crucial starting point by generating initial visuals. 
Teacher 1, who participated in Phases 1-3, noted that DALL-E provided a good starting point for students with deficiencies in drawing or a lack of imagination, \textit{``Project-based courses can be difficult for students, and they often need step-by-step guidance on creating artworks. 
It can be challenging for students who are not skilled in drawing or designing to express their ideas in artwork. DALL-E helped them initiate the creative process and express themselves more freely in their artwork}''.
As for GPT, teachers highlighted its value in improving students' efficiency in completing projects. 
They remarked that students could search for all the necessary information within a single interface and receive tailored answers to their specific questions, thereby ``\textit{shortening the time spent on information gathering}``. 
Teacher 1 pointed out in Phase 3, ``\textit{students needed to come up with words related to both dragons and the college entrance exam as well as finish the drawing of a poster within one 40-minute course session. Time was very limited, but with GenAIs, they succeeded}''.

\penguin{\textbf{Finding 11: Teachers concern that GenAIs may reduce learning outcomes if students use them inappropriately.}}
Teachers expressed concerns about the potential for GenAIs to undermine learning outcomes when students fail to use them appropriately. Teacher 2 emphasized the importance of helping students effectively incorporate GenAIs into their regular PBL activities, highlighting the need for both teachers and students to learn how to balance their time and effort when interacting with these tools. She noted, ``\textit{both us and students need time to figure out how to effectively integrate GenAIs into their regular art courses and manage their time and effort interacting with GenAIs.}''
Teachers also emphasized that the appropriateness of using GenAIs depends on the nature of the art project and the students' abilities. For example, they were more open to GenAI use in tasks requiring information seeking and solution ideation, such as handcraft projects in Phases 1 and 2 or poster designs in Phases 3 and 4. However, Teacher 1 stressed that students’ skill levels should inform how GenAIs are used. She explained, ``\textit{If this group of students are not good at drawing and crafting objects, it is acceptable that they directly mimic the content generated based on their ideas, which can train their basic art skills. However, if the group is proficient in drawing and crafting objects, it is not acceptable that they copy the generated content in their artworks, as I would expect that they can enact critical and creative thinking to come up with unique artworks.}''
Additionally, teachers expressed concern about projects that could be easily completed by directly using GenAI-generated content without further processing. For instance, if students relied solely on AI-generated materials to appreciate a painting or replicate an object, the use of GenAIs might bypass opportunities for learning. Teacher 1 suggested that students should submit the AI-generated materials they referenced along with their final artwork, explaining, ``\textit{I can assess whether they simply copy the materials.}''

\section{Discussion}\label{sec:discussion}
\pzh{
Based on our findings, this section discusses implications for the usage of GenAIs in group project-based learning activities, teachers' roles in GenAI-enhanced art courses, and future GenAI-powered learning support tools for elementary students. 
}

\ReviseZQW{
\subsection{Elementary Students’ Use of Generative AIs in PBL}
Based on the findings of RQ1 about elementary students' usage of GenAIs in a Chinese PBL art course, our study provides insights into the opportunities and challenges in these interactions. Firstly, our findings demonstrate that elementary school students can understand and effectively use GenAI tools. Despite their developmental stage and limited exposure to advanced technologies \cite{van2023emerging}, students intuitively leveraged the tools’ distinct affordances in a sequential manner, using GPT for information seeking and DALL-E for visual inspiration (Finding 2). This structured approach highlights the students’ ability to adapt to the unique strengths of these tools. GPT served as a textual guide, helping students address knowledge gaps and explore creative possibilities, while DALL-E supported visual ideation by translating abstract concepts into tangible representations. Importantly, this sequential usage was not explicitly taught; students independently aligned each tool with specific stages of their project workflows. This intuitive integration showcases the accessibility of generative AI tools for elementary learners, countering concerns about their potential opacity or complexity {\ReviseFinal{\cite{vader}}}.

Moreover, GenAI tools were proved effective in addressing the challenges elementary students face during the testing phase of PBL. Evidenced by significantly higher satisfaction levels reported by students using \name{} compared to the baseline group, GPT provided support for project information seeking and idea generation(Finding 1), bridging gaps in students' knowledge and sparking creativity. DALL-E complemented this by enabling students to visualize and refine abstract ideas (Finding 1), facilitating the translation of concepts into tangible representations. Together, these tools reduced the challenges often encountered during the ideation and implementation phases of group projects, which are frequently attributed to students' lack of knowledge and skills in teamwork and self-directed learning \cite{chen2021forms, miao2024project}. Additionally, by offering immediate and individualized support, generative AIs acted as a valuable resource in classrooms where teachers were unable to provide real-time assistance to all groups {\ReviseFinal{\cite{giannakos2024promise}}}, enhancing students’ ability to progress independently in their projects.

However, the challenge of formulating queries(Finding 3) highlights a critical gap in the accessibility of generative AI tools for younger users. While prompting challenges also occur in higher education settings, where users may struggle with optimizing specificity or creativity \cite{palani2021conotate}, the difficulties faced by elementary school students are fundamentally rooted in their developmental stages. Younger learners often lack the computational thinking skills and familiarity with AI systems required to align their goals with the affordances of these tools \cite{lee2021developing}. This gap limits their ability to navigate interfaces effectively and to grasp the iterative nature of prompting, leading to a greater reliance on external assistance, as observed in our findings (Finding 3), and a reliance on trial-and-error approaches.

As for groups using AI, the two collaboration strategies(Finding 4) observed in the student groups—assigning fixed members versus allowing everyone to interact with \name{}—also reflect different ways students navigated group dynamics. 
{\ReviseFinal{Fixed-member strategies facilitated structured collaboration by designating a specialist to mediate AI interactions. This approach introduced a built-in process for aligning the group on challenges and solutions, as the specialist gathered input from the team and reported AI-generated insights. However, it also centralized control over GenAI interactions, risking having personal bias for conversations with GenAI. On the other hand, allowing all students to interact freely with GenAI provided greater flexibility, enabling individuals to explore AI-generated suggestions independently. Yet, this approach did not necessarily enhance group engagement, particularly for students who were already less involved in collaboration. While the introduction of GenAI influenced collaboration dynamics, team engagement ratings (\autoref{fig:findings1} (b)) showed no significant difference between the experimental and baseline groups. This suggests that while GenAI offers opportunities to support collaboration—such as by facilitating consensus-building \cite{lin2024text}—its use alone does not inherently lead to improved teamwork or engagement without deliberate scaffolding and coordination \cite{han2024teams}.}}
}
% Fixed-member strategies likely offered efficiency, enabling groups to designate a "specialist" for interacting with the tools. On the other hand, shared interaction encouraged inclusivity {\ReviseFinal{\cite{han2024teams}}}, allowing all members to engage with the technology and contribute to the creative process. Both strategies come with trade-offs: while fixed roles may streamline tasks, they could limit participation, whereas shared use promotes broader engagement but requires more coordination. These findings echo prior research on collaborative learning, which emphasizes the importance of balancing structure and participation in group work \cite{isohatala2020cognitive,jeong2016seven}.  
\ReviseZQW{
\subsection{Design Considerations for GenAI-Powered Learning Tools Targeting K-6 Students}}
\ReviseZQW{Based on the findings of RQ2, we propose three key considerations for designing GenAI-powered learning tools tailored to K-6 students. }

\subsubsection{Streamlining the input flow to GenAIs based on students' usage patterns}
\Original{
our findings revealed that students often used GPT and DALL-E in a sequential order\ReviseZQW{(Finding 2)}. 
\peng{\name{} streamlines this flow with a ``Select and Generate'' option (\autoref{fig:askart2} A5), which enables students to select specific words or sentences from the chat with GPT and use them as the main subject or context words to generate images in DALL-E}. 
\ReviseZQW{It was proved to be helpful for help students iterate their queries to DALL-E(Finding 8).}
These findings are in line with previous research \peng{\cite{jiang2023graphologue}} showing that a structured input feature can assist users in constructing effective prompts to GenAIs.
Therefore, we suggest that future GenAI-powered learning support tools should equip features that make it easy for elementary students to have subsequent queries to GenAIs in their learning activities. 
These tools can be inspired by previous work on chaining large language model (LLM) prompts, which has been shown {\ReviseFinal{to improve}} the quality of outcome and sense of human-AI collaboration \peng{\cite{wu2022ai}}. 
For example, based on the user pattern of iteratively refining queries to DALL-E observed in our study, future tools could suggest prompts \eg about related objects, backgrounds, and goals, to improve the generated images in the last turn. Nevertheless, developers of these learning support tools should observe students' usage patterns in the target learning activities and iteratively improve their tools, which is a requirement in the user-centered design process.}

\ReviseZQW{\subsubsection{Guiding Initial Use and Sustained Engagement with GenAIs}
In Phases 3 and 4, we observed that providing context-specific example queries significantly improved students’ understanding and proficiency in using GenAIs (Finding 5). For instance, the project-related self-introduction of \name{} and the example queries embedded in its interface (\autoref{fig:askart1} A2) were frequently adopted by students to initiate interactions. These use cases, preset by the research team to align with the specific themes and tasks of each course, enabled students to engage with GenAIs more confidently and effectively. Extending this feature, future tools could offer customizable interfaces for teachers to design example queries tailored to their classroom context. Templates based on project themes, student skill levels, and learning goals could be integrated with LLM capabilities to automatically generate relevant use cases, further supporting students in their initial interactions with GenAIs.

Beyond familiarity, keeping students engaged during conversations with GenAIs remains critical. The “Suggested Follow-Ups” feature in \name{} (Finding 7) demonstrated its potential to transform how students perceive and interact with GPT. {\ReviseFinal{We observed that elementary students tended to approach GPT with a one-turn query mindset, similar to how they might use a traditional search engine,}} missing the opportunity to leverage the conversational depth and iterative power unique to GenAIs. The suggested follow-ups helped introduce students to the concept of multi-turn dialogue, guiding them to {\ReviseFinal{build on previous responses and ask further questions}}. This shift not only facilitated deeper {\ReviseFinal{engagement, but also allowed students to explore}} more sophisticated, layered thinking through their interactions.

\subsubsection{Scaffolding Query Articulation for Elementary Students}

While the “Audio-to-Text Input” feature aimed to address challenges in typing and query formulation (Finding 3), it proved to be ineffective (Finding 6). Students’ voice inputs were often disorganized or incomplete, resulting in unsatisfying responses from GPT. This underscores a more fundamental challenge: elementary students frequently struggle to clearly and correctly express their questions. Addressing this unique challenge requires solutions that go beyond simply mitigating typing difficulties. Future tools should focus on scaffolding students’ ability to articulate their queries, starting with features that help them structure and clarify their thoughts. For example, tools could provide guided templates{\ReviseFinal{\cite{jamplate, di2022idea}}}  or visual prompts{\ReviseFinal{\cite{son2024genquery, sain2021stylemeup}}} that break down complex queries into simpler components, helping students focus on their intent and the necessary details. {\ReviseFinal{Additionally, timely and user-adaptive support can proactively detect ambiguities or incomplete queries and engage users in interactive clarification and refinement\cite{zhang2023clarify, erbacher2022interactive}}}. By tackling the root of these challenges, such tools could empower students to engage in more meaningful and productive interactions with GenAIs, ultimately enhancing their learning experience and fostering greater independence in their use of AI systems.

}

\subsection{\ReviseZQW{Teacher Roles in Leveraging GenAI for Effective PBL Classrooms}}
Based on the findings of RQ3, our findings indicate three considerations for teachers' roles in GenAI-enhanced PBL art courses. 
% In line with previous studies\cite{robinson2024reviewing}, our findings reflect that teachers are already aware of their pivotal role in AI-driven PBL classes. K-6 students represent a special group of users for GenAI, as these tools can be difficult to comprehend due to their complex and distributed nature\cite{van2023emerging}. Therefore, teachers should give extra consideration throughout PBL classes.

\ReviseZQW{
\subsubsection{Customizing GenAI Tools to Align with PBL Stages}
Teachers’ support for integrating GenAIs into PBL from the study (Finding 9) highlights the potential of these tools to enhance inquiry-driven learning. Beyond simply deploying GenAIs for information seeking and solution ideation, teachers could play an active role in customizing the settings of these tools to align with the specific stages of the PBL process. Customizations could include project-specific contexts, themes, or structured templates that guide students in crafting queries and reflecting on AI outputs. 
For instance, teachers could design prompts or configurations in GPT that encourage students to explore targeted concepts, themes, or methods related to their projects, enabling focused and guided exploration. Similarly, DALL-E settings could be adjusted to emphasize visual outputs that align with particular styles, techniques, or cultural elements central to the learning objectives. These adjustments ensure that students’ interactions with GenAIs remain both relevant and immersive, providing a structured yet flexible framework for exploration and ideation. This approach aligns with prior work that highlights how structured frameworks can help novices organize information, think critically, and improve creative ideas by embedding LLM capabilities into familiar templates\cite{jamplate}.

\subsubsection{Adapting GenAI Guidance to Student Abilities and Task Complexity}Teachers should tailor their guidance on GenAI usage to address the varying abilities of students while aligning it with the nature of the PBL task. Teachers noted that DALL-E provided a strong starting point for students with deficiencies in drawing or imagination, while GPT improved efficiency by organizing information and supporting task completion (Finding 10). These strengths make GenAIs particularly effective for abstract or complex tasks requiring extensive contextual understanding. For example, teachers could introduce pre-generated suggestions or scaffolded prompts to help less confident students engage with GenAIs, while advanced students could refine outputs or explore iterative design tasks requiring greater creative freedom.
For simpler tasks—such as creating posters in Phase 4—conventional tools like search engines may suffice, allowing students to focus on creative ideation without the cognitive load of navigating GenAIs. In Phase 4, no significant difference was observed in perceived challenge levels between the experimental group (AskArt) and the control group, and the baseline group reported slightly higher creativity scores (\autoref{fig:findings1} (b)). However, for conceptually challenging projects, integrating GPT for synthesizing information and DALL-E for visual ideation can deepen inquiry and creativity. By aligning the use of GenAI with the complexity of tasks and student needs, teachers can foster inclusive collaboration while enhancing the PBL workflow.}

% Third, 
% teachers should build up an evaluation plan for the usage of GenAIs in students' art projects. 
% The increasing use of AI by students in learning has made it challenging to assess learning outcomes in PBL \peng{\cite{zheng2024charting}}. 
% Our teachers suggested that students should submit the usage record of GenAIs along with the required artwork of the projects. 
% As reflected by the teachers in our study, the acceptability of students referencing GenAI-generated text/images depends on the type of artwork (whether it can be directly generated by GenAI) and the students' capabilities. 
% They deemed it acceptable for a student lacking artistic skills to closely mimic a DALL-E image if they have provided detailed instructions and demonstrated their creative thinking. 
% To guide students document their usage of GenAIs, \citet{zheng2024charting} present multiple designs that could be usable, \eg regarding the task allocation between students and AI and the process of students incorporating AI's suggestions into the project. 
% \peng{Nevertheless, teachers need to keep in mind that the requirements of documenting AI usage should be as simple as possible, such that the students can finish the submission of their projects within a course session.}

\ReviseZQW{
\subsubsection{Aligning GenAI Use with Task Nature and Student Abilities to Reduce Misuse}
Teachers should ensure that students’ use of GenAIs aligns with PBL learning objectives by addressing concerns about inappropriate use(Finding 11), such as over-reliance on AI outputs or superficial engagement . We recommend that GenAI integration consider the nature of the task and students’ abilities. For tasks involving information seeking and ideation, GenAIs are highly effective, but students with stronger skills should be encouraged to go beyond replication and demonstrate originality and critical thinking. Conversely, students with limited skills might use AI-generated content as a foundation to develop their abilities. To address these concerns, teachers can require students to document their GenAI interactions, including prompts, outputs, and their decision-making processes. This aligns with \citet{zheng2024charting}, who advocate for tracking task allocation and integration of AI suggestions to provide insight into student contributions. For tasks that could be completed easily with AI outputs, teachers might also require students to submit AI-generated materials alongside their final work to evaluate how they engaged with and adapted the outputs.
}

\subsection{Limitations}
\pzh{Our work has several limitations.
First, since the teaching assistants (TA) were not presented in the course sessions in Phase 1, and there were no recordings or audio captures, all qualitative feedback in Phase 1 was derived from the interviews with the teacher. 
This lack of comprehensive resources might have resulted in overlooking some potential design requirements for \name{}.  
Second, in Phase 4, no preliminary materials about GenAI were provided to the students, and they were given only 40 minutes to learn and use it. This short interaction time may have led to incomplete feedback from the students.
Finally, from Phase 2 to Phase 4, the involvement of three TAs may create an environment that differs from a regular classroom setting in elementary schools. 
In the future, when students get familiar with the usage of GenAI, we could observe the students' behaviors without the presence of TAs in more art course sessions. \ReviseZQW{Finally, our experimental design did not include a comprehensive baseline comparison. A more systematic approach, as conducted by prior work\cite{ye2024contemporary}, would involve equipping baseline groups with computer interfaces that allow the use of search engines for broader data collection. }
%a longer-term field study (\eg 10 weeks) could be implemented. 
% Fourth, the questionnaire responses from elementary students exhibit strong subjectivity and some randomness. For example, in Phase 4, some students who did not get the chance to use GenAI may have given lower ratings because they were unhappy with the arrangements. Additionally, despite providing small prizes to encourage careful and thoughtful responses, some students might have quickly marked all maximum or minimum scores without consideration. The TAs interviewed 5 of them who gave extreme scores, and they all provided reasonable explanations. Therefore, these scores were not filtered out during the analysis.
}

\Original{
\section{Conclusion}
In this study, we investigated the integration of GenAI in K-6 project-based art courses through a four-phase field study. Our research revealed the multifaceted impact of GenAIs, highlighting both their benefits and challenges. Students experienced enhanced information, inspiration, and guidance for their projects but struggled with query formulation to GenAI. The introduction of \name{}, a tailored interface combining DALL-E and ChatGPT, facilitated improved interaction patterns and increased engagement, yet indicated areas needing further refinement. Different collaboration strategies emerged among students, with increased engagement observed due to GenAIs. Teachers appreciated the enhanced student participation but voiced concerns about potential misuse and the need for more suitable interfaces for elementary school students. Based on the findings, our study further discussed key areas including information seeking, team collaboration, and the evolving roles of teachers in GenAI-enhanced classrooms. This research offers valuable insights into the effective use of GenAIs in K-6 project-based learning, presents \name{} as a practical solution for customizing GenAI interactions, and provides actionable recommendations for educators and researchers. These contributions aim to support the development and implementation of GenAI-powered learning tools in elementary education, advancing our understanding of their potential to transform student learning and engagement.}

  \bibliographystyle{ACM-Reference-Format}
  \bibliography{mybibliography}

\end{document}